\documentclass[11pt]{article}

\usepackage[T1]{fontenc}
\usepackage[utf8]{inputenc}
\usepackage{mathptmx}
\usepackage{CJKutf8}
\usepackage[margin=1in]{geometry}
\usepackage{setspace}
\usepackage{graphicx}
\usepackage{booktabs}
\usepackage{tabularx}
\usepackage{array}
\usepackage[labelfont=bf,labelsep=period,font=small]{caption}
\usepackage{hanging}
\usepackage{xurl}
\usepackage[hidelinks]{hyperref}
\usepackage{authblk}

\title{\textbf{How K--12 Teachers Adapt Generative AI to Local Classroom Demands: Cultural Distance Between GenAI Defaults and Everyday Teaching Practices}}

\author[1*]{Qing Xiao}
\author[1*]{Ruiwei Xiao}
\author[2]{Hanqi Jane Li}
\author[1]{Phenyo Phemelo Moletsane}
\author[3\textdagger]{Xinying Hou}
\author[1\textdagger]{John Stamper}
\author[1\textdagger]{Hong Shen}
\affil[1]{Human-Computer Interaction Institute, Kenneth C. Griffin School of Computer Science, Carnegie Mellon University, Pittsburgh, Pennsylvania, USA}
\affil[2]{Department of Education Studies, University of California, San Diego, La Jolla, California, USA}
\affil[3]{Department of Data Science, College of Computing, New Jersey Institute of Technology, Newark, New Jersey, USA}
\date{\small Accepted for publication in {\textit{International Journal of Human--Computer Interaction}}}

\begin{document}

\maketitle

\renewcommand{\thefootnote}{}
\footnotetext{* Qing Xiao and Ruiwei Xiao contributed equally to this work as co-first authors. \textdagger\ Xinying Hou, John Stamper, and Hong Shen contributed equally to this work as co-senior authors.}
\footnotetext{Corresponding authors: Qing Xiao (\href{mailto:qingx@andrew.cmu.edu}{qingx@andrew.cmu.edu}) and Ruiwei Xiao (\href{mailto:ruiweix@cs.cmu.edu}{ruiweix@cs.cmu.edu}), Human-Computer Interaction Institute, Carnegie Mellon University, 5000 Forbes Avenue, Pittsburgh, PA 15213, USA.}
\renewcommand{\thefootnote}{\arabic{footnote}}

\begin{abstract}
Generative AI (GenAI) is entering K--12 classrooms, but its default culture does not always fit local teaching. Existing accounts either treat this misfit as macro-level structural bias or examine it through single-case studies; neither captures how much work teachers must do to make outputs usable across tasks. This exploratory, theory-building qualitative study introduces cultural distance, the gap between GenAI's default cultural repertoire and the situated demands of teachers' practice. Through interviews with 30 K--12 teachers who frequently use GenAI, 10 each from South Africa, Taiwan, and the United States, we identified three levels defined by the effort teachers invested and whether it yielded a usable output: low (minor edits), mid (sustained prompting and revision), and high (structural gaps rooted in unsupported languages, undigitized knowledge, or policy filters that no teacher effort can close). We present cultural distance as a promising framework requiring further examination and draw implications for designers, policymakers, and educators.

\end{abstract}

\section{Introduction}

Generative AI (GenAI) is rapidly entering classrooms worldwide, offering teachers tools for everyday work (Cordero \& Cordero-Castillo, 2025; Memon \& Kwan, 2025). These GenAI systems are attractive because they can rapidly generate materials that support teachers' core responsibilities, such as lesson preparation (Karpouzis et al., 2024), simplifying complex concepts for students (Elnaem et al., 2025), and classroom communication (Hirata \& Hirata, 2025). However, despite these benefits, GenAI is not culturally neutral. Trained on large-scale, culturally uneven datasets (Liu, 2023), GenAI applications like OpenAI ChatGPT (OpenAI, 2022), Google Gemini (Google DeepMind, 2023), or ByteDance Doubao embody a ``default culture'' that privileges norms and knowledge from certain dominant cultures, often those of their countries of origin, while marginalizing other cultural perspectives (Tao et al., 2024). While this orientation usually enables outputs to align easily with teachers' expectations in routine tasks, such as summarizing meeting reports, mismatches emerge when local curricula, linguistic diversity, or culturally specific pedagogical practices are absent. In such cases, teachers must localize these culturally default outputs from GenAI: sometimes with minor edits, sometimes through extensive reworking, and sometimes finding adaptation impossible when, for example, Indigenous languages are not supported (Mhasakar et al., 2025). Inspired by these emerging phenomena, in this paper, we focus on \textit{how K--12 teachers use chat-based GenAI (like ChatGPT) to support their everyday teaching practices and how they experience, adapt to, or struggle with GenAI's outputs and local classroom demands}.

Existing accounts of this misfit sit at two levels. Macro-level critiques treat it as structural bias in how AI is built, at the extreme as ``AI colonialism'' (Hao, 2022). Micro-level work describes how individual users negotiate it in use, by refining prompts, reframing tasks, or rejecting outputs (Fan et al., 2025), or documents particular teachers and tools one case at a time. Neither level captures what teachers actually encounter across tasks. What is missing is a middle-range analytic lens (Merton, 1968) between the two that captures the spectrum of alignment teachers meet in practice, ranging from seamless integration of GenAI, to adaptation requiring substantial cultural labor (reprompting, editing, and supplementing outputs with the teacher's own cultural knowledge), to cases where its limitations make adaptation impossible. By a middle-range lens we mean an approach situated at the level of teachers' professional practices across contexts. It abstracts from individual teachers' localized efforts to reveal recurring patterns of cultural distance in GenAI adoption, patterns that are neither as abstract as global critiques of ``AI colonialism'' nor as narrow as single-case accounts.

To address this gap, we first introduce the concept of cultural distance. We define cultural distance as \textit{the gap between GenAI's default cultural repertoire and the situated demands of teachers' practice}. By culture, we adopt a multi-layered view that encompasses not only national or regional traditions, but also the micro educational contexts, including the specific curricular standards, teachers' own pedagogical philosophies, the languages teachers and students use in class, and communicative norms that shape everyday teaching.

We then develop this framework of cultural distance through an exploratory, theory-building qualitative study based on 30 in-depth interviews with K--12 teachers in South Africa, Taiwan, and the United States. The aim of the study is to develop and refine the concept, not to characterize K--12 teachers in general. These contexts were deliberately chosen through a maximum-variation logic, differing in linguistic diversity, educational traditions, how much AI training data exists in their languages, and the presence or absence of domestic AI development. Teachers described using GenAI across various educational contexts, including communication, lesson design, assessment, and culturally responsive teaching, with each aspect illustrating different levels of cultural distance and corresponding efforts to bridge them.

Our analysis revealed six recurring forms of cultural distance between GenAI defaults and local classroom demands. (1) At the low-distance level, GenAI outputs are often aligned smoothly with teachers' needs, for example, in drafting communication with parents and students (\textit{L-1}) or brainstorming engaging instructional activities (\textit{L-2}). (2) At the mid-distance level, teachers faced partial misalignments that required significant adaptation: assessment questions pitched at the wrong difficulty, which teachers fixed through repeated prompting and editing (M-1), and vague treatment of local cultures, which teachers supplemented with their own cultural knowledge (M-2). (3) At the high-distance level, teachers encountered structural barriers that lay beyond what teacher effort alone could close, for instance, when local languages and traditions were missing from training data (\textit{H-1}) or when queries were blocked outright by policy restrictions (\textit{H-2}). These findings show that cultural distance is not binary but exists on a spectrum. In some cases, GenAI fits easily, in others it demands heavy cultural labor, and in still others the gap is structural, lying beyond what any teacher effort can bridge.

This study makes three contributions to education, Human-Computer Interaction (HCI), and the broader AI community: (1) Theoretically, it introduces a middle-range concept that links GenAI model defaults with local teaching practices, moving beyond both abstract critiques of bias and narrow case studies of practice. We conceptualized teacher-driven alignment as a spectrum of relations between GenAI defaults and local scenarios. Unlike accounts that make teachers' adaptation labor visible, the concept makes that labor comparable across tasks, tools, and regions, and it holds in one frame the cases where no adaptation is needed, where adaptation is effortful but possible, and where no effort can close the gap. (2) Empirically, it provides an exploratory account of how teachers in three deliberately varied contexts adopt and adapt GenAI in everyday practice, identifying both shared strategies (prompt refinement, editing, reframing) to align their needs with GenAI outcomes and context-specific challenges (language exclusions, curricular misalignments, policy restrictions) that hinder such teacher-driven alignment. (3) Practically, the framework offers a vocabulary for researchers to analyze \textit{AI alignment} as a spectrum. It provides policymakers with insights into how infrastructure and policy can narrow or widen the distance. It also makes visible to teachers the cultural labor involved in adapting AI tools after adoption.

\section{Literature Review}

Research on AI in education has highlighted both how K--12 teachers are beginning to integrate AI into their teaching practices and factors that shape the conditions and pathways of teachers' AI adoption. A third line of research examines the role of culture, showing how local norms, languages, and institutional contexts shape both technology adoption and classroom practices. However, these strands share a blind spot. Studies of teachers' AI adoption largely stop at whether and why teachers take up these tools, and work on culture tends to prescribe sensitivity as a principle; neither asks how closely GenAI's outputs actually sit to the situated demands of classroom practice, or how much effort teachers must spend to close that gap. We propose \textit{cultural distance} as a lens for exactly this question.

\subsection{K--12 Teachers' AI Adoption for Teaching Practice}

AI is rapidly becoming intertwined with K--12 education, and teachers around the world are experiencing and responding to this trend (Bernstein, 2025; Chounta et al., 2022; Filiz et al., 2025; Zhang \& Tur, 2024). While much existing research has discussed its integration into K-12 student learning (Kazemitabaar et al., 2023; Klar, 2025; D. Wang et al., 2025) and assessment (Cohn et al., 2024; Hou et al., 2025), recent HCI research has also investigated this topic across different K-12 contexts from the teacher's perspective.

For example, Ahn and Lim (2025) reported that physical education teachers valued AI to help them address operational challenges in K-12 physical education classes by serving as an operational assistant and student data evaluator. In special education, Lin et al. (2025) investigated 52 special education teachers and found that while AI tools can deliver high-quality and diverse content, they failed in deeply personalized accommodations. What's more, Hedderich et al. (2024) explored how middle school teachers designed chatbots to assist their teaching about cyberbullying safely. A design workshop with K-12 teachers from the United States also highlighted how AI can help them to support students in programming learning, e.g., diagnosing help-seeking and giving process-focused feedback (Limke et al., 2025). However, AI can also create challenges for teachers, such as increasing workload and reducing autonomy, as highlighted by Harvey et al. (2025) from a multi-stakeholder perspective. Relatedly, Yin et al. (2026) examined how K--12 teachers anticipate and respond to breakdowns when classroom AI systems conflict with their pedagogical judgments, characterizing teachers as engaging in ongoing ``pedagogical repair'' to sustain Responsible AI values such as fairness, accountability, and relational care. In a U.S. district that was an early adopter of GenAI, Dangol et al. (2026) found that teachers neither simply adopted nor resisted the tools but negotiated their role, welcoming relief from routine tasks while worrying about what would be displaced, and that this negotiation rested on hidden interpretive labor, such as crafting prompts, reviewing outputs critically, and adapting them to students' needs.

Despite these insights, prior work tends to examine AI use in K--12 education at the level of individual tools within single-region contexts (Ahn \& Lim, 2025; Dangol et al., 2026; Hedderich et al., 2024; Lin et al., 2025). How teachers across \textit{different cultural regions} experience and adopt GenAI in divergent ways and what conceptual lens might explain such variation remains largely unexamined.

\subsection{Factors Influencing Teachers' Adoption of AI}

Not all teachers worldwide have adopted AI at the same pace or in the same manner. Multiple factors might influence teachers' AI adoption practices. From an individual level, teachers' perceived usefulness and AI literacy level can influence their behavioral intention to adopt AI (S. Ma \& Lei, 2024). Teachers with stronger self-efficacy and AI understanding report more benefits, fewer concerns, and greater trust in AI (Viberg et al., 2025). Beyond individual-level differences, prior work has also shown that school-level factors (technological and cultural) can also influence teachers' behavioral intention to use AI (Wu et al., 2025). A survey across U.S. states found that, while teachers generally have positive attitudes and openness to AI, gaps in technology access and comfort level with technology persist across regions, genders, and age groups (Woodruff et al., 2023). From the technology side, smooth AI tool adoption is expected to prevent extra workload, provide sufficient support, and address teacher ownership and ethics (Cukurova et al., 2023).

While these studies explain \textit{what makes teachers more or less likely to adopt} AI, they largely stop at the moment of adoption. Far less attention has been paid to what happens afterward, namely the ongoing work teachers must do to make a generally capable tool actually fit their classrooms once it is in use. We take up this post-adoption labor directly, proposing \textit{cultural distance} as a middle-range framework that links structural imbalances in AI design to teachers' everyday practices.

\subsection{Cultural Sensitivity and Localizations of AI in Education}

It is important to recognize that culture profoundly shapes both classroom teaching and the adoption of technology in classrooms. Being culturally sensitive in this context means planning, using, and evaluating AI/EdTech from the local classroom outward, starting with teachers' and students' languages, histories, and values, and ensuring that classroom power dynamics are respected so the tool fits the class rather than forcing the class to fit the tool (Plata, 2008). Research in parent education, teachers' professional development, and discipline programs shows that culturally sensitive approaches can increase teacher confidence, respect community norms, and improve uptake across diverse populations (Awad et al., 2016; Smith et al., 2017; J. Wang et al., 2025). For AI specifically, this implies designing classroom AI tools, and the professional development programs that introduce them, to adapt to local values and contexts, rather than assuming that teachers everywhere will take up the same tools in the same way.

Beyond sensitivity as a stance, a longstanding tradition in education theorizes culture as a central pedagogical resource. Culturally relevant pedagogy (Ladson-Billings, 1995) argues that effective teaching builds on students' cultural references to develop academic success and cultural competence; culturally responsive teaching (Gay, 2018) extends this by using students' cultural knowledge and frames of reference to make learning more meaningful. More recently, culturally sustaining pedagogy (Paris, 2012) reframes the goal from bridging students toward a dominant norm to actively sustaining their languages, literacies, and community practices. Across these accounts, culture shapes not only how content is delivered but what counts as legitimate knowledge in the classroom, positioning curriculum, student identity, and community knowledge as core teaching resources. And it raises the stakes for GenAI, since a tool that misrepresents or omits these cultural resources is not merely less convenient but pedagogically consequential.

Culture is also multi-layered. National culture is not the only layer that matters; professional and disciplinary cultures interact with country effects (Nistor et al., 2013). Even within a single country, teachers' experiences are far from uniform. For example, Hawaiian kumu report cultural misalignment and low-resource language gaps as real limits, even when these AI tools could save time (Mhasakar et al., 2025); in North American classrooms, researchers found that the interaction patterns and GenAI outputs can vary by learners' characteristics such as first-language (B. Ma, 2024) or self-efficacy (R. Xiao et al., 2024).

Despite these insights, much of this work treats cultural sensitivity as a design \textit{principle} to be advocated, rather than offering a way to \textit{analyze} how closely AI outputs actually sit to teachers' situated practices. What is missing is not another call for sensitivity but an analytic handle on \textit{fit}, one that registers not only whether a tool is culturally attuned, but how much work teachers must do, from a quick edit to no workable path at all, to make it usable. Cultural distance offers such a handle, linking structural asymmetries in data and policy to the lived adaptations of teachers in South Africa, Taiwan, and the United States.

\section{Proposing the Cultural Distance Framework}

This section develops \textit{cultural distance} as the conceptual backbone of our study. We proceed in four steps. We first review why existing accounts of GenAI's cultural misfit, though valuable, do not give us what we need. We then define cultural distance and position it against the established construct of the same name in international business. Next, we describe the three-phase process through which the concept took shape, so that our later claims can be read against the work that produced them. Finally, we delimit the concept's scope and explain what it is designed to make visible.

\subsection{What Structural and Alignment Accounts Leave Out: The Degree of User Effort}

Prior HCI and AI research provides essential background for understanding why GenAI often misaligns with local classroom practices. From the natural language processing (NLP) perspective, dataset studies have shown that languages and regions are unequally represented in training corpora (Bender et al., 2021; Joshi et al., 2020), and benchmark analyses confirm that these imbalances lead to systematic disparities in model performance (Li et al., 2025; Lyu et al., 2025). Such disparities extend beyond low-resource languages. For example, the Natural Questions dataset, as one of the most widely used English-language corpora, contains over 80\% U.S.-sourced content (Faisal et al., 2022), embedding strong regional skews. Because textual data carry cultural assumptions, models trained predominantly on U.S.-centric corpora implicitly adopt American norms as the ``default'' (Lu et al., 2025). For teachers elsewhere, this default translates into extra cultural work, such as revising, rewriting, or supplementing GenAI outputs that presuppose Americanized classroom practices.

HCI research similarly shows that producing culturally localized systems is costly and labor-intensive, often requiring ethnographic investigation and sustained adaptation (Höök, 2000; Reinecke \& Bernstein, 2011, 2013; Yeo, 1996). Economic structures exacerbate this imbalance, as countries with higher GDPs are more likely to advance domestic AI research and products (Blasi et al., 2022), while global markets remain dominated by U.S., Chinese, and European firms (Maslej et al., 2025). As a result, teachers in lower-resource regions are more likely to face outputs misaligned with their classrooms. They must expend additional effort to adapt global GenAI tools, which are not designed with their specific contexts in mind.

Beyond structural explanations, recent work on value alignment emphasizes that misalignment is not resolved once at the level of model design but is continually negotiated in everyday use. Fan et al. (2025) describe this process as user-driven value alignment, where people actively reshape AI behavior through strategies such as refining prompts, reframing tasks, or rejecting outputs they view as harmful. Users attempt to steer AI toward outcomes that better reflect their own values and contexts. Shen, Knearem, et al. (2024) and Shen, Ma, et al. (2024) extend this line of work with the notion of bidirectional alignment, emphasizing that alignment is a two-way process. Humans are expected to develop new literacies for working with agentic AI, such as learning how to collaborate effectively. In turn, AI systems must be designed to absorb human feedback and embed social values. Read together, these studies shift attention from purely developer-side fixes to the situated work users perform during interaction. What they do not yet offer, however, is a way to compare \textit{how much} of this work different tasks demand, for instance, whether a given misalignment is closed by a quick edit or resists adaptation entirely.

Collectively, these strands illuminate why and how misalignment arises. Structural accounts (data, infrastructure, markets) explain why GenAI's defaults sit far from many classrooms in the first place, while value-alignment research shows that the resulting gaps are negotiated in everyday use. What neither offers is a middle-range lens for the \textit{degree} of user work this negotiation requires, that is, a way to say not just \textit{that} a teacher must adapt an output, but \textit{how far} they must travel to make it usable, and whether the journey is even possible. This gap motivates the concept we introduce next.

\subsection{Defining Cultural Distance in the AI-In-Education (AIED) Context}

To address this gap, we propose \textit{cultural distance}, defined as \textit{the gap between GenAI's default cultural repertoire and the situated demands of teachers' practice}. By culture, we refer not only to national or regional traditions, but also to curricular standards, pedagogical philosophies, the languages teachers and students use in class, and communicative norms that shape everyday teaching. In this sense, GenAI's ``default cultural repertoire'' includes not only representational patterns in training data but also the product policies, safety filters, and institutional assumptions through which certain topics, languages, or pedagogical uses become supported or constrained. Accordingly, we do not use cultural distance as a catch-all label for every mismatch between GenAI outputs and classroom needs. Rather, the concept applies when such mismatches involve local knowledge, language, pedagogical expectations, communicative norms, or culturally and institutionally situated assumptions about what counts as appropriate classroom content and use. General technical failures, hallucinations unrelated to local context, latency, or interface usability problems fall outside this scope unless they are tied to these cultural or institutional dimensions.

The word ``default'' in this definition needs stating precisely, because a language model does not give one fixed answer. Its output changes with how the question is asked, with what was said earlier in the conversation, and with the user's own wording, and models trained on human feedback tend to shift their answers toward what the user seems to want. By a model's default we mean what it produces before the user has done any of this work: the answer to a plain request, before reprompting, giving examples, or editing by hand. That answers can be improved by effort is the starting point of the concept, since the amount of effort needed is what cultural distance measures. The limit is that effort can only move an answer within what the model's training data, fine-tuning, and product rules allow it to produce. A model's willingness to agree with the user does not extend this limit: it can adopt a teacher's framing in its wording while still lacking the knowledge to act on it. The default repertoire, then, is this whole range of answers a model can be brought to give, together with the answer it starts from when no one has pushed. Low and mid distance describe how far a teacher must push within the range; high distance describes a need that falls outside it.

The word \textit{''distance,''} for its part, is not new. It originates in international business and cross-cultural management research, where Kogut and Singh (1988), building on Hofstede's (1980) value dimensions, formalized it as a composite index of the degree to which national cultures differ from one another, and where it has since been widely used to predict firms' foreign-market entry, entry-mode choice, and affiliate performance (Shenkar, 2001; Sousa \& Bradley, 2008). We borrow the core insight of \textit{distance} from this tradition, namely the intuition that cultural difference can be treated as a graded rather than binary quantity. We do not, however, adopt its measurement apparatus or its national-level assumptions. Where the international-business index measures a fixed, symmetric distance between two national cultures, our notion is directional and enacted. It runs from a globally trained artifact's defaults toward the situated demands of a particular classroom, and becomes observable only through the effort a teacher must expend, varying by task and setting. In treating distance as something continually negotiated in use, our framing aligns with Shenkar's (2001) call to attend to the \textit{interface} between interacting cultures and the \textit{friction} their contact produces, rather than to an abstract gap between them.

A natural alternative to distance is proximity, cultural or linguistic, which names the same relation from the other end. The two terms direct attention differently. Proximity names a property of the fit between a system and a context; distance names what must be traversed to reach fit. Because our central claim is that this relation becomes observable only through the effort teachers expend, and because our contribution is to make that labor comparable across tasks, we retain the term that names the traversal. Distance also keeps the concept anchored to the international-business tradition from which we borrow its graded character, and it accommodates the high end of our spectrum more naturally, as a gap that no effort can close is readily described as a distance that cannot be covered, whereas ``zero proximity'' does not convey that the obstacle is structural rather than a matter of degree. Proximity, and in particular linguistic proximity, remains a productive complementary framing for work that asks how close GenAI already sits to a given language, curriculum, or community, and we return to it as a direction for future research in the Limitations and Future Work section.

\subsection{How the Concept Was Developed}

\textit{Cultural distance} did not arrive ready-made; it emerged through a three-phase, iterative process in which an initial framing drawn from prior work was sharpened in conversation with practitioners and then refined against empirical data. We describe these phases here so that the conceptual claims we make later can be read against the process that produced them. Figure 1 summarizes the three phases, what each took as input, and what each produced.

\begin{figure}[htbp]
\centering
\includegraphics[width=\textwidth]{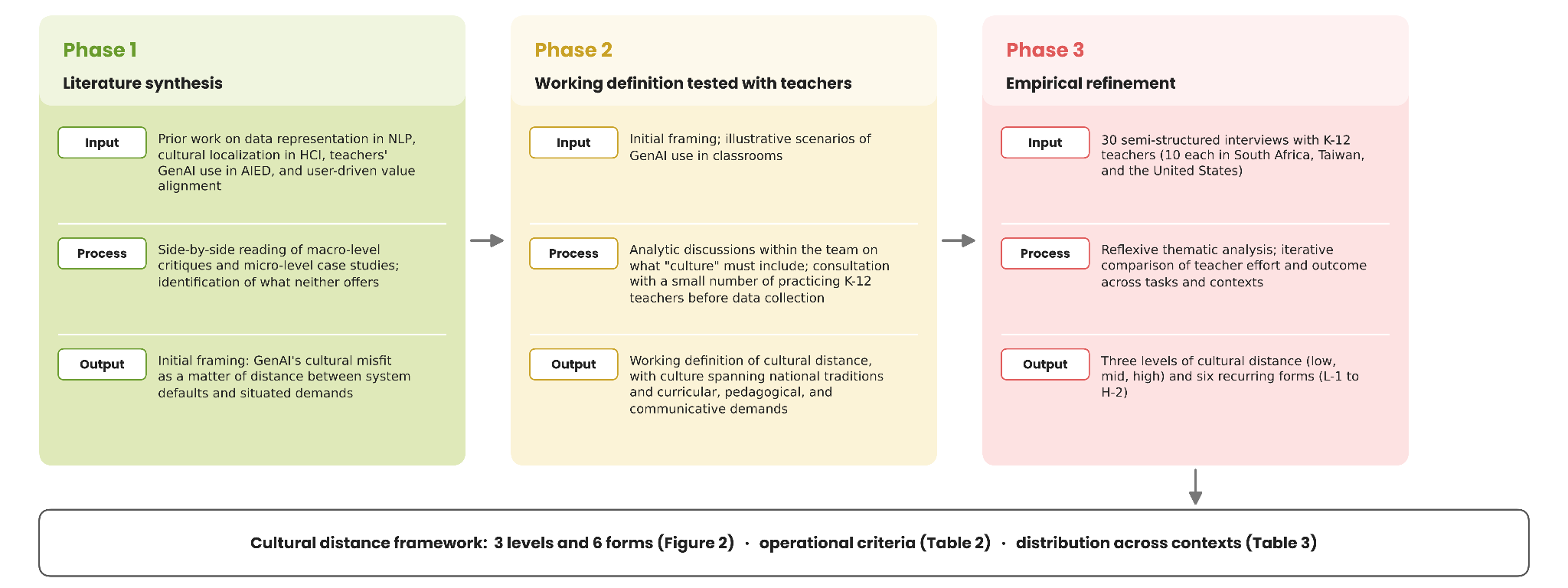}
\caption{The Three-Phase Process Through Which the Cultural Distance Framework Was Developed}
\label{fig:1}
\end{figure}

\textbf{Phase one: literature synthesis.} The first phase was the literature synthesis presented above. Reading research on data representation in NLP, the cost of cultural localization in HCI, work on AI in education (AIED) documenting how K--12 teachers adopt and adapt GenAI in their everyday practice, and user-driven and bidirectional value alignment side by side, we noticed that accounts of GenAI's cultural misfit clustered at two extremes: macro-level critiques of structural bias or ``AI colonialism,'' and micro-level case studies of particular tools in particular settings. What neither extreme offered was a lens for the \textit{degree} of work global users must perform to make a generally capable system usable in a specific context. This absence, more than any single finding, motivated our initial framing of the phenomenon as a matter of \textit{distance} between a system's default cultural repertoire and the situated demands of practice.

\textbf{Phase two: a working definition, tested with teachers.} In the second phase, the research team worked to turn this framing into a workable definition. Through a series of analytic discussions, we debated what \textit{culture} should encompass in this context and concluded that a narrowly national or linguistic reading would be too coarse. Teachers' difficulties arose as much from curricular standards, pedagogical philosophies, grade-level expectations, and communicative norms as from language or nationality. We therefore defined culture broadly, spanning both national and regional traditions and these more micro educational contexts, and articulated the core intuition that the same tool could sit near or far from a teacher's needs depending on the task and setting.

To avoid fixing the concept solely from the researchers' vantage point, we then shared this working definition, together with illustrative scenarios, with a small number of practicing K--12 teachers before formal data collection began. We asked whether the notion of distance matched their own experience of using GenAI, where it fit, and where it broke down. Their responses served three purposes. They confirmed the concept's face validity, indicating that teachers recognized the tensions we were trying to name. They helped us broaden and sharpen its scope, for instance by treating unsupported local languages and policy restrictions as part of the same spectrum. And they refined our wording so that distance clearly denoted the variable effort different tasks demanded.

\textbf{Phase three: empirical refinement.} In the third phase, we refined the concept's internal structure empirically. The 30 semi-structured interviews and the reflexive thematic analysis that followed (see the Methods section) gave the framework its dimensions. We allowed teachers' accounts to specify the levels, which surfaced the spectrum of low, mid, and high cultural distance and the six recurring categories within it (L-1 through H-2). The analysis also fed back into the concept itself. For example, comparing cases across regions clarified the boundary between mid-distance situations, where sustained effort eventually closes the gap, and high-distance situations, where it does not. In this sense, the literature and early consultation produced the concept, while the empirical material gave it shape and tested its limits.

To be clear, the concept of cultural distance was neither induced from the interview data alone nor imposed on the data as a fully specified framework in advance. Rather, we used cultural distance as an initial sensitizing concept developed through prior literature and early practitioner consultation (Phases one and two). The interview data then specified its empirical structure (Phase three). Through reflexive thematic analysis of teachers' accounts of adaptation and usability, we identified the low-, mid-, and high-distance levels, as well as the six recurring forms within those levels.

\subsection{Scope and What the Concept Makes Visible}

Building on this process, we refine and ground the framework in the sections that follow. Teachers' accounts of everyday classroom practice across South Africa, Taiwan, and the United States provide the lived experiences through which cultural distance becomes observable, allowing us to develop the low-to-high spectrum and demonstrate its analytic value.

We develop this concept through teachers' everyday engagements with chat-based GenAI tools that support their teaching practices. Unlike technical perspectives that focus on APIs or model-level interventions, our lens centers on how frontline teachers interact with GenAI through natural language dialogue in routine classroom tasks, such as drafting emails, designing activities, or generating assessment questions, and how they make these outputs usable through their own efforts and adaptations.

Yet, as the notion of \textit{cultural distance} suggests and as our later findings sections detail, these efforts do not always guarantee success. At times, minor adjustments are enough to make GenAI outputs effective; in other cases, teachers must expend substantial time and creativity to achieve only partial alignment; and in high-distance situations, even intensive effort cannot bridge the gap, leaving the outputs unusable. This variability underscores why cultural distance is a useful analytic concept. It draws attention not only to teachers' adaptations but also to the limits of what their efforts can accomplish when global defaults fail to represent local classroom demands.

\section{Methods}

We conducted an exploratory, theory-building qualitative study based on 30 in-depth, semi-structured interviews with K--12 teachers who were actively integrating GenAI into their teaching practices. As an interview study, it produces teachers' first-person accounts rather than experimental measurements, and the Findings section accordingly presents these accounts, organized by the level and form of cultural distance they illustrate. To capture a gradient of cultural distance, we recruited 10 teachers each from South Africa, Taiwan, and the United States, contexts that differ in cultural background and in how much AI training data exists in their languages. Each interview lasted 60--90 minutes and was conducted remotely via Zoom. Participants received USD 20 per hour in Amazon gift cards as compensation. Our inquiry focused on how teachers described their experiences of adopting GenAI products into the classroom, including both challenges and opportunities. These accounts informed our empirical refinement of cultural distance, helped identify its underlying forms, and guided us in developing design implications.

\subsection{Context: AI Adoption in Education in South Africa, Taiwan, and the United States}

In this section, we provide an overview of the socio-economic, technical, and cultural contexts shaping teachers' adoption of AI in the three regions represented in our study. These contextual elements lay the foundation for situating our findings and interpreting their broader significance. Each context contains far more variation than ten teachers can capture, across provinces, school types, and language communities; we use the three contexts as a source of contrast for developing the concept.

\subsubsection{South Africa}

South Africa presents a distinctive educational context shaped by both deep socio-economic inequality and a high unemployment rate, as well as linguistic diversity. With one of the world's highest Gini indices (Chitiga et al., 2014; Spaull \& Jansen, 2019; Statista, 2025), colonial and apartheid legacies have left wealth concentrated in the hands of a minority (Seekings \& Nattrass, 2005). The socio-economic disparities are mirrored in the linguistic landscape. South Africa recognizes 12 official languages, including the historically dominant languages English and Afrikaans, widely spoken Indigenous languages such as Zulu and Xhosa (Alexander, 2025), and South African Sign Language (Ethnologue, 2025). With the exception of English, South Africa's official languages remain low-resource for AI. In addition, locally developed AI products are scarce. As a result, most users rely on U.S.-developed English-language tools (BusinessTech, 2024). The dominance of English and Afrikaans extends into the educational system. From Grade 4 onward, about 80\% of learners are taught in English or Afrikaans, although most do not speak either language at home (Howie et al., 2008), and by 2024, over 90\% of the 23,719 public schools used English as the main medium, with roughly 10\% using Afrikaans exclusively or alongside English (Al Jazeera, 2024).

\subsubsection{Taiwan}

Taiwan preserves rich cultural diversity. It includes high-tech regions like the Hsinchu Science Park (often dubbed ``Taiwan's Silicon Valley'') and vibrant Indigenous communities (2.5\% of the total population; Council of Indigenous Peoples, 2024), whose oral traditions, rituals, and ancestral practices (such as hunting and weaving) are rarely documented in digital form (Tu, 2025). From a natural language processing (NLP) perspective, while Mandarin Chinese is broadly recognized as a high-resource language in NLP (Joshi et al., 2020), this categorization does not distinguish between its two written systems: Traditional Chinese (used in Taiwan) and Simplified Chinese (used in mainland China). Divergences between these systems, including variations in vocabulary and usage, cross-strait cultural and ideological differences, and a substantial imbalance in model training corpora, could pose challenges for adapting large language models to Taiwan's educational and social contexts (Li et al., 2025; Lyu et al., 2025).

\subsubsection{United States}

The U.S. is taking a leading role not only in AI development broadly (Maslej et al., 2025) but also in the educational technology market, contributing the largest revenue share of 38\% in 2024 (Grand View Research, 2025). Major generative AI applications such as ChatGPT (OpenAI, 2022) and Gemini (Google DeepMind, 2023), as well as widely used teacher-facing AI tools like MagicSchool (MagicSchool AI, 2023) and Khanmigo (Khan Academy, 2023), are all U.S.-developed, making the landscape of AI tool use highly U.S.-centric. This advantage can be attributed to, and further reinforced by, the status of English as a high-resource language, which dominates training corpora and contributes to disproportionate progress in language model development (Joshi et al., 2020; Maslej et al., 2025). Regarding teacher adoption in the U.S. education system, 60\% of U.S. K--12 teachers reported using AI tools in their work during the 2024--25 school year, with 32\% indicating weekly use and an average time savings of 5.9 hours per week (Gallup, 2025).

\subsection{Participants: K--12 Teachers Using GenAI Tools in Their Everyday Work (N=30)}

We sought to recruit as diverse a sample as possible across the three regions in order to capture variations in cultural distance and to ground our theoretical framework in empirical evidence. Inclusion criteria required that participants were active K--12 teachers with direct classroom responsibilities, were frequent users of GenAI in their teaching, and showed interest or experience in its cultural dimensions and localization. We recruited 10 teachers per context, a size chosen to give depth within each setting and contrast across settings for an exploratory, theory-building study; the aim was to develop the concept, not to estimate the prevalence of its forms.

Table 1 summarizes the demographic and professional characteristics of the 30 teachers we interviewed; its last column records which of the six forms of cultural distance reported in the Findings section each teacher's account contributed to. Across the three regions, our sample reflected diverse representation in terms of gender, years of teaching experience, school types, and subjects taught. The gender ratio was consistently 7:3 (female to male) in all three regions, with the majority of teachers working in public schools. In terms of GenAI tools, ChatGPT was the most widely used across all contexts. Meta AI was also frequently mentioned in South Africa, mainly due to its integration into the widely used WhatsApp platform in the country. In addition, education-specific GenAI tools such as MagicSchool were reported by teachers in both the United States and South Africa.

\begin{table}[htbp]
\centering
\scriptsize\setlength{\tabcolsep}{4pt}
\setstretch{1.0}
\caption{Participant Demographics}
\label{tab:1}
\begin{tabularx}{\textwidth}{@{}l l >{\raggedright\arraybackslash}p{1.3cm} >{\raggedright\arraybackslash}X >{\raggedright\arraybackslash}X >{\raggedright\arraybackslash}X >{\raggedright\arraybackslash}p{1.9cm}@{}}
\toprule
\textbf{\textbf{ID}} & \textbf{\textbf{Gender}} & \textbf{\textbf{Years of Experience}} & \textbf{\textbf{School Type}} & \textbf{\textbf{Teaching Subject}} & \textbf{\textbf{GenAI Tools Used}} & \textbf{\textbf{Cultural Distance Mentioned}} \\
\midrule
S1 & Male & 0--5 & Public High School & English, Sepedi (Local Language) & ChatGPT, Meta AI, Grok & H-1 \\
S2 & Female & 0--5 & Public High School & Math, English & ElevenLabs & L-2, M-2 \\
S3 & Female & 11--15 & Public High School & English, Afrikaans (Local Language) & ChatGPT, Meta AI & L-2, H-1 \\
S4 & Female & 0--5 & Public Elementary School, High School & Math, Science, English, Physical Education & ChatGPT & L-2, H-1 \\
S5 & Female & 0--5 & Public High School & Business, Life Orientation & ChatGPT, Meta AI & H-1 \\
S6 & Female & 0--5 & Public High School & English, Economics, Tourism & Gemini, Meta AI & L-1, M-1, H-1 \\
S7 & Male & 11--15 & Private High School & Math, Life Sciences & ChatGPT, Gemini, Grok & L-2 \\
S8 & Female & 6--10 & Private High School & Science, Chemistry & ChatGPT, MagicSchool, Mindjoy & L-1, L-2, M-1, M-2, H-1 \\
S9 & Female & 0--5 & Public Preschool, Elementary School & Math, English, Life Orientation & ChatGPT & L-2, H-1 \\
S10 & Male & 0--5 & Public Elementary School & English, Social Sciences, Life Orientation & ChatGPT, Grok & L-1, L-2, M-1 \\
U1 & Male & 6--10 & Public High School & Social Science & Gemini & L-2 \\
U2 & Female & 11--15 & Public High School & English & ChatGPT, Gemini, MagicSchool & L-2, M-1 \\
U3 & Female & 20+ & Public High School & Statistics, CS, Math & ChatGPT, Gemini & L-2 \\
U4 & Male & 11--15 & Public High School & English, Creative Writing & ChatGPT & H-2 \\
U5 & Female & 0--5 & Public Elementary School & Special Education & Gemini & L-1, L-2, M-1 \\
U6 & Male & 20+ & Public High School & Social Studies, History, Civics & ChatGPT, Gemini & L-1, L-2, M-1 \\
U7 & Female & 6--10 & Public High School & English & ChatGPT & L-1 \\
U8 & Female & 16--20 & Public High School & English, Social Studies, Economics & ChatGPT, Gemini, MagicSchool & L-1, M-1, M-2 \\
U9 & Female & 11--15 & Private Elementary School & Science, Social Studies & ChatGPT, MagicSchool & L-1 \\
U10 & Female & 20+ & Public High School & English, History & ChatGPT, Gemini & L-1 \\
T1 & Female & 0--5 & Public Middle School & Information Technology & ChatGPT & L-2 \\
T2 & Male & 20+ & Public Elementary School & Information Technology, Arts & ChatGPT, Doubao (\begin{CJK}{UTF8}{bsmi}豆包\end{CJK}), ERNIE Bot (\begin{CJK}{UTF8}{bsmi}文心一言\end{CJK}) & M-2, H-2 \\
T3 & Female & 0--5 & Public Kindergarten & Life Skills & ChatGPT & L-2 \\
T4 & Female & 6--10 & Public Elementary School & Special Education & ChatGPT & L-1, M-1 \\
T5 & Female & 20+ & Public High School & Information Technology & ChatGPT & L-2 \\
T6 & Female & 11--15 & Public Middle School & Information Technology & ChatGPT & L-2, M-1, M-2 \\
T7 & Male & 20+ & Public High School & Information Technology & ChatGPT, Gemini, NotebookLM & L-2 \\
T8 & Male & 6--10 & Private High School & STEM & ChatGPT & L-2 \\
T9 & Female & 0--5 & Public Elementary School & Special Education & ChatGPT & L-2, M-1 \\
T10 & Female & 0--5 & Public Kindergarten & Life Skills & ChatGPT, Copilot & L-2, M-2, H-1 \\
\bottomrule
\end{tabularx}
\par\smallskip\raggedright \textit{Note.} S = South Africa; U = United States; T = Taiwan. Codes in the last column refer to the six forms of cultural distance (L-1 to H-2) identified in the findings.
\end{table}

\subsection{Semi-Structured Interviews}

This study was approved by the IRB (Institutional Review Board). We conducted semi-structured interviews to explore how K--12 teachers use GenAI in their daily work, with particular attention to moments where cultural distance emerged between system outputs and local classroom practices. Each interview lasted between 60 and 90 minutes and was conducted remotely via Zoom.

The interview protocol was organized around four topics related to cultural distance. First, we asked teachers to describe their everyday use of GenAI in teaching, including specific tasks (e.g., drafting emails, preparing lesson activities, generating assessments) and the motivations that led them to experiment with these tools. Second, we probed challenges, frustrations, or mismatches they encountered, focusing on how outputs aligned or failed to align with their students' learning levels, local curricula, or community expectations. Here, teachers were encouraged to reflect on their emotional responses, ranging from relief and confidence to stress, disappointment, or even distrust.

Third, we investigated the concrete efforts teachers made to adapt GenAI outputs. This included prompting strategies, manual revisions, supplementation with their own cultural or pedagogical knowledge, or switching to alternative platforms designed for education. These accounts allowed us to capture the uneven labor teachers performed in negotiating cultural distance.

Finally, we asked explicitly about local culture and practices. Teachers reflected on how language, regional or national curriculum standards, Indigenous or community traditions, and broader cultural values shaped their expectations of GenAI, as well as their sense of whether the outputs felt ``usable'' in their contexts. We also asked how school-level policies or state regulations influenced their ability to adopt or adapt GenAI.

This structure enabled us to capture not only what teachers did with GenAI, but also how they felt about the results, what additional work was required to make outputs useful, and how local cultural factors shaped their perceptions of alignment and misalignment.

\subsection{Data Analysis}

Our analysis was guided by the concept of \textit{cultural distance}, developed through the three-phase process described in the ``Proposing the Cultural Distance Framework'' section. At the beginning of analysis, however, we did not treat low, mid, and high cultural distance as predefined categories. Rather, we used cultural distance as an initial sensitizing concept: a working definition that directed our attention to mismatches between GenAI's default cultural repertoire and the situated demands of teachers' classroom practice. The internal structure of the framework, including the three levels of cultural distance and the six recurring forms reported in the findings, was developed iteratively through engagement with the interview data.

We employed reflexive thematic analysis (Braun \& Clarke, 2019) to examine how teachers described the fit between GenAI outputs and their classroom needs, the effort required to make outputs usable, and the points at which such effort succeeded or failed. We first coded transcripts for concrete practices, such as drafting emails, designing activities, generating assessments, and preparing culturally relevant materials; teachers' motivations and frustrations; the kinds of mismatches they encountered; and the strategies they used to adapt, revise, or abandon GenAI outputs. At this stage, our codes remained close to participants' accounts; the low/mid/high structure came later.

In subsequent rounds of analysis, we compared these codes across cases and clustered them according to recurring patterns of teacher effort and outcome. Some tasks required only minor editing before GenAI outputs became usable; others required sustained prompting, revision, or supplementation with teachers' own pedagogical and cultural knowledge; still others remained unusable despite teachers' attempts to adapt them. Through this iterative comparison, we developed the three-level spectrum of low, mid, and high cultural distance. We then refined the six more specific categories within this spectrum by examining which classroom tasks, cultural demands, and adaptation strategies recurred across participants.

Table 2 summarizes the criteria that resulted from this process. Two features of an account determined its level: the kind of effort the teacher reported, and whether that effort yielded an output the teacher could use. The levels are therefore not separated by a numeric threshold but by qualitative markers in what teachers described doing and what happened as a result. The boundary between mid and high distance was the one most often at issue; we assigned an account to high distance only when the teacher reported having tried more than one strategy and still obtaining no usable output, or when the obstacle lay in a product policy or institutional rule that no prompting strategy could address. Accounts in which a teacher reported that sustained effort eventually produced a usable output were assigned to mid distance even when the effort was considerable. When coders initially disagreed about whether an account was mid or high, the case was resolved in discussion by checking whether the teacher reported a usable output at the end, and, where no usable output resulted, whether the cause lay in an unsupported language, undigitized local knowledge, or a policy filter. Accounts that on inspection involved only generic failures, such as hallucinations unrelated to local context, latency, or interface problems, were set aside as outside the scope of cultural distance.

\begin{table}[htbp]
\centering
\small
\setstretch{1.0}
\caption{Operational Criteria for the Three Levels of Cultural Distance}
\label{tab:2}
\begin{tabularx}{\textwidth}{@{}>{\raggedright\arraybackslash}p{1.0cm} >{\raggedright\arraybackslash}X >{\raggedright\arraybackslash}X >{\raggedright\arraybackslash}X >{\raggedright\arraybackslash}p{1.2cm}@{}}
\toprule
\textbf{\textbf{Level}} & \textbf{\textbf{Effort reported by the teacher}} & \textbf{\textbf{Outcome}} & \textbf{\textbf{Indicative markers in accounts}} & \textbf{\textbf{Forms}} \\
\midrule
Low & Minor edits to wording, tone, or format; a single prompt usually sufficed & Usable output, often exceeding expectations & ''I just changed a few words''; output adopted almost directly & L-1, L-2 \\
\addlinespace
Mid & Sustained prompting, manual revision, supplementing with own pedagogical or cultural knowledge, or switching to an education-specific tool & Usable output, but only after considerable effort & Repeated reprompting; ''I have to change the details all by my own''; use of textbook content or local references in prompts & M-1, M-2 \\
\addlinespace
High & Multiple strategies attempted, or no strategy available & No usable output; task abandoned or done without GenAI & Wrong or refused responses in local languages; hallucinated local knowledge; queries blocked by product filters or institutional bans; reversion to English or to non-AI methods & H-1, H-2 \\
\bottomrule
\end{tabularx}
\end{table}

To support analytic rigor and reflexivity, the coding process involved three researchers with experience in AI education research and qualitative methods. These researchers first independently reviewed a subset of transcripts and met regularly to compare interpretations, refine codes, and discuss how teachers' accounts related to the emerging concept of cultural distance. Our goal was to use collaborative coding and discussion to surface multiple interpretations and sharpen the analytic distinctions among different forms of fit, effort, and breakdown. When ambiguity concerned whether a mismatch reflected a situated cultural, linguistic, or institutional demand rather than a generic technical failure, the team discussed these cases with research members who had long-term research or teaching experience in the relevant region, including South Africa, Taiwan, and the United States. These discussions helped us avoid treating unfamiliar local practices as generic mismatches and instead examine whether they reflected situated demands around language, curriculum, pedagogy, communicative norms, or legitimate classroom knowledge. In this sense, our team's varied regional, educational, and methodological experiences served as an important reflexive resource for interpreting cultural distance across contexts.

Although our participants came from South Africa, Taiwan, and the United States, we did not set out to produce a region-by-region comparison. Instead, we treated cross-regional diversity as a methodological resource for sharpening the concept of cultural distance. By including teachers who worked under different educational systems, languages, and policy environments, we were able to capture a wider range of GenAI use cases and identify variation in how cultural distance manifested across settings. Similar patterns of fit, effort, and breakdown appeared across regions, but their specific forms differed: for example, assessment questions could be oversimplified in U.S. classrooms, misaligned with learner levels in Taiwan, or require textbook-specific contextualization in South Africa; high-distance cases could involve unsupported local languages, Indigenous knowledge, or policy-level restrictions. These cross-regional patterns helped us distinguish the levels and forms of cultural distance. They suggest, but do not establish, that the framework may apply beyond any single regional context; establishing that is a task for future work with larger and more diverse samples.

\begin{figure}[htbp]
\centering
\includegraphics[width=\textwidth]{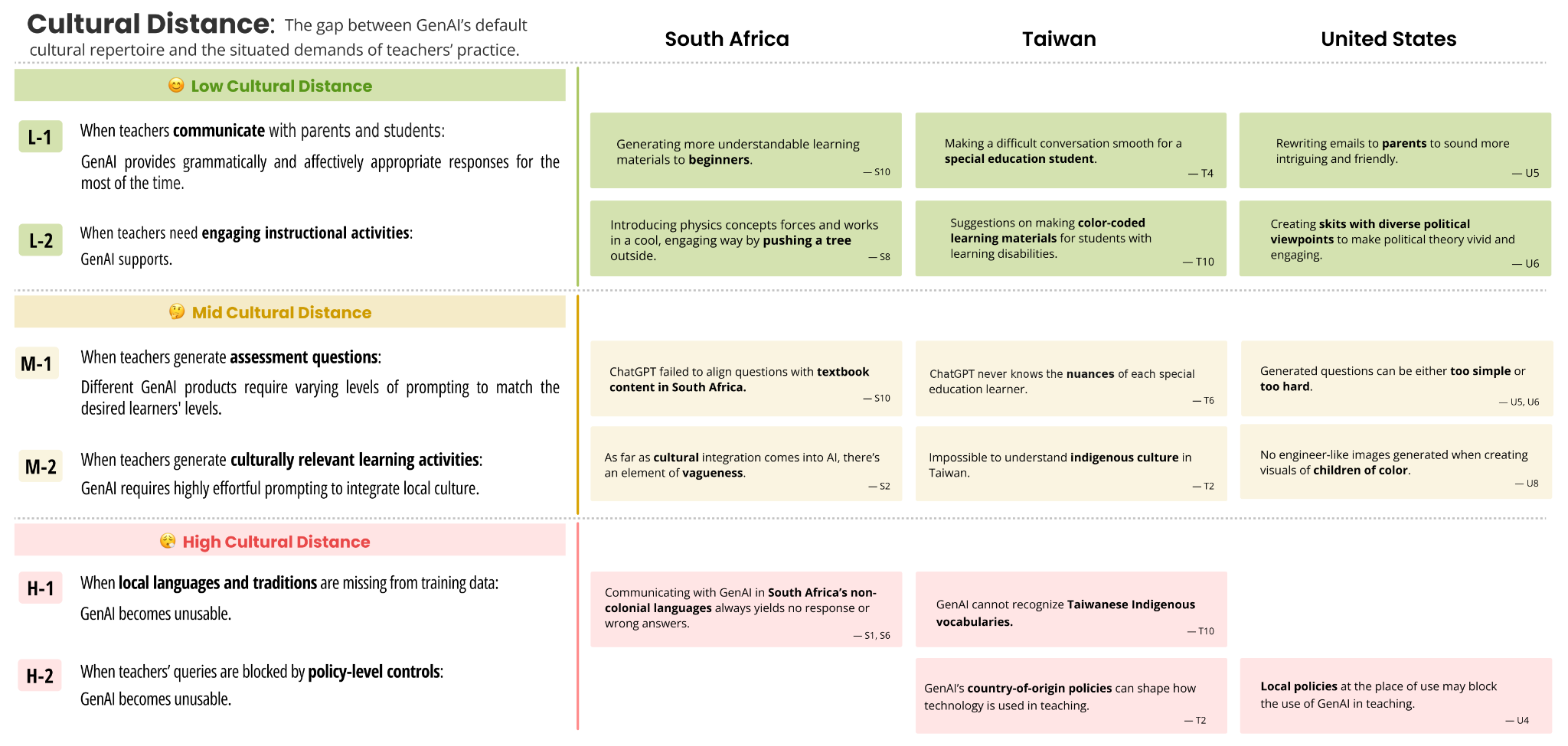}
\caption{A Spectrum of Cultural Distance in Teachers' Experiences With GenAI}
\label{fig:2}
\par\smallskip{\small\raggedright \textit{Note.} Teachers' experiences with GenAI can be understood along a spectrum of \textit{cultural distance}: the gap between GenAI's default cultural repertoire and the situated demands of teaching practice. We identified three levels of cultural distance, each defined by the varying amount of effort teachers must invest when using GenAI to support their teaching practices, ranging from low to high. Within each level, we identified two distinct categories, for a total of six. For every category at each cultural distance level, we included one teaching example from each of the three cultural contexts (South Africa, Taiwan, and the United States), where available.\par}
\end{figure}

\section{Findings}

Our analyses show that teachers' experiences with GenAI can be understood along a spectrum of cultural distance: the gap between GenAI's default cultural repertoire and the situated demands of teaching practice. Figure 2 summarizes the framework that resulted: its rows are the six forms of cultural distance grouped into three levels, its columns are the three contexts, and each cell gives an example from that context where one was available, so that the figure shows at a glance both the structure of the framework and the evidence behind each form. We identified three levels of cultural distance, each marked by different amounts of effort required from teachers when using GenAI to support teaching practices: (1) At the low-distance level, GenAI outputs generally met expectations in routine tasks such as communication with parents and students or generating engaging activities, requiring only minimal adjustment (see the Low Cultural Distance subsection). (2) At the mid-distance level, teachers encountered partial mismatches, for example, when creating assessments or culturally relevant activities, which demanded significant prompting and revision (see the Mid Cultural Distance subsection). (3) At the high-distance level, teachers faced structural barriers such as unsupported local languages or policy restrictions, where adaptation was either extremely difficult or altogether impossible (see the High Cultural Distance subsection). These three levels illustrate how alignment with GenAI varies across tasks and contexts, shaping the degree of cultural work teachers must perform to make AI usable in their classrooms. The six forms we describe were identified in the accounts of 30 teachers who were frequent GenAI users; we present them as recurring patterns within this sample, and other forms may appear with other teachers and in other settings. Table 3 shows how the 30 teachers' accounts were distributed across the six forms and the three contexts.

\begin{table}[htbp]
\centering
\small
\setstretch{1.0}
\caption{Number of Teachers Whose Accounts Contributed to Each Form of Cultural Distance, by Context}
\label{tab:3}
\begin{tabularx}{\textwidth}{@{}>{\raggedright\arraybackslash}X c c c c@{}}
\toprule
\textbf{\textbf{Form}} & \textbf{\textbf{South Africa (n = 10)}} & \textbf{\textbf{Taiwan (n = 10)}} & \textbf{\textbf{United States (n = 10)}} & \textbf{\textbf{Total}} \\
\midrule
L-1 Communicating with parents and students & 3 & 1 & 6 & 10 \\
L-2 Designing engaging instructional activities & 7 & 8 & 5 & 20 \\
M-1 Generating assessment questions & 3 & 3 & 4 & 10 \\
M-2 Generating culturally relevant learning activities & 2 & 3 & 1 & 6 \\
H-1 Local languages and traditions missing from training data & 7 & 1 & 0 & 8 \\
H-2 Queries blocked by policy-level controls & 0 & 1 & 1 & 2 \\
\bottomrule
\end{tabularx}
\end{table}

\subsection{Low Cultural Distance: Little Effort from K-12 Teachers to Get Quality GenAI Outcomes for Teaching}

This section highlights two aspects where the cultural distance was low to K--12 teachers: using GenAI for communication tasks (see L-1) and for brainstorming engaging instructional activities (see L-2). In both cases, the cultural demands of these tasks (the register expected in school--home communication and the conventions of what makes a classroom activity engaging) fell squarely within the cultural territory that GenAI's default repertoire most densely represents. The low distance we report is therefore a matter of overlap rather than of these tasks being culture-free: the norms teachers needed to hit were the norms the model had most thoroughly absorbed. As a result, teachers could adopt GenAI outputs with only minimal editing, yet still reported outcomes that often exceeded their expectations. These low-distance cases reveal how even small amounts of effort produced disproportionately high value, reducing stress in communication and sparking new energy in classroom activities.

\subsubsection{When Teachers Communicate with Parents and Students, Most of the Time GenAI Provides Grammatically and Affectively Appropriate Responses}

Communication is a central yet demanding part of teachers' daily work, involving frequent emails, notices, and sensitive conversations with parents and students. Teachers in our study emphasized that effective communication requires not only grammatical accuracy but also the right emotional tone, which can be challenging under time pressure or when confidence is low. K-12 teachers expected that GenAI might ease this burden by supporting the dual demands of fluency and empathy in communication.

In practice, GenAI often exceeded these expectations in communication support for our participants. Teachers described how it generated usable messages with very minimal adjustments, highlighting the relatively low cultural distance in this domain.

For example, T4, a special education teacher in Taiwan, described feeling ``blank'' when asked to inform a parent about their child's transfer and not knowing how to phrase the message. She turned to GenAI for support, which produced step-by-step advice sensitive to both the autistic child and the parents, in T4's view: \textit{``The child is autistic, with high anxiety and high sensitivity. And my brain was blanked, I thought to myself, how would I know the best way to talk about this? So I asked ChatGPT, and it suggested: you could first take the child to walk around the new school, let him play there a few times, and gradually highlight the positive aspects of this new environment.''}

T4 recalled being surprised not just by the fluency of the phrasing but by the specificity of the advice, which felt culturally sensitive to both the child and the parents. Later, she emphasized that the outcomes surpassed her own expectations after following GenAI's suggestions: \textit{``The child did not react with the kind of severe emotional distress we had expected from a change of environment. Instead, he accepted it rather quickly... the whole process was very peaceful, smooth, and successful.''}

This sense that GenAI often outperformed teachers' own communication abilities was echoed across contexts. S10 valued how it helped him \textit{``switch-code concepts from the textbook into English that beginners could actually understand''}, noting that such instant simplifications had previously required much more time. U5 similarly used GenAI to \textit{``rewrite this email so that it sounds more intriguing or so that it sounds more friendly''}, describing how the tool not only made her messages clearer but also gave her greater confidence.

These accounts show that communication tasks represent a case of \textbf{low cultural distance (L-1)}. Because general-purpose GenAI models are trained to produce fluent and emotionally attuned text, their ``default culture'' is already aligned closely with teachers' needs. As a result, teachers only needed minor edits, yet often obtained phrasing and tone that surpassed what they could have produced under stress. Among all teaching-related activities, this was where participants reported the most significant help: school--home interactions became smoother, emotional burdens were reduced, and the payoff from GenAI felt disproportionately high compared to the minimal effort required.

\subsubsection{When Teachers Need Engaging Instructional Activities, GenAI Supports}

Designing class activities that capture and sustain students' attention is a universal challenge for K--12 teachers, but one where GenAI outputs aligned strikingly well with classroom needs. Because the engagement conventions teachers sought (dramatization, role-play, and vivid framing) are widely represented across the educational materials GenAI is trained on, the model's defaults sat close to what teachers needed. The distance here was minimal because these particular conventions are densely encoded in the default repertoire. Teachers across regions consistently reported that GenAI suggestions could be adopted almost directly, requiring minor adjustment yet producing high levels of classroom engagement.

S8, a South African science teacher, expected help in making abstract concepts like \textit{force} and \textit{work} accessible and engaging. She turned to the tool Mindjoy, which suggested a hands-on warm-up: \textit{``Take a walk with them, and then go around the tree. Let them try to push a tree as hard as they can, and then ask, `What do you think is happening?'''} The system further guided her on introducing ``work'': \textit{``Now, if I want to introduce work done, is work actually being done there? No---because the object did not move, even though you applied a force.''} Reflecting afterwards, S8 felt GenAI enabled her to \textit{``introduce forces in a cool way,''} and her students found the activity \textit{``so fun, like, so engaging.''}

U6, a U.S. humanities teacher, expected new ways to make political theory more vivid. He asked GenAI to create skits dramatizing different perspectives, and the tool produced named characters with libertarian, conservative, and liberal viewpoints. He explained: \textit{``The AI creates names and viewpoints for each character, and then students read the skit---it really helps highlight what I'm teaching.''} The outcome, he noted, was that students engaged more deeply with abstract political concepts than when he explained them directly.

U5, an elementary special education teacher, expected variety in classroom storytelling, since students had grown restless hearing the same tales. She used GenAI to generate new narrated stories for grades 1--3, commenting: \textit{``They just can't listen to the Three Little Pigs for the 15th time.''} The new stories, delivered in distinct voices, kept children engaged while freeing her to monitor classroom behavior, an outcome she found more effective than repeating familiar materials.

S2, a South African English teacher, wanted to bring Shakespeare to life in ways textbooks could not. She used ElevenLabs to render a soliloquy in Shakespeare's own \textit{``voice.''} In her words: \textit{``It makes it look like Shakespeare was in the room with us. Learners forget they are actually learning, and they just start learning either way.''} The attempt exceeded her expectation, as students not only paid attention but also became immersed in the performance.

Finally, T10 expected more visually stimulating materials for students with learning disabilities. She adopted GenAI's suggestions on how to \textit{``make slides and learning materials more colorful,''} and found this directly improved students' ability to maintain focus during lessons.

These accounts show why we classify activity design as a \textbf{low cultural distance (L-2)} task. Teachers' expectations centered on generating engaging activities, and GenAI's default repertoire already overlapped with these needs. With minimal effort to adapt their outputs, they achieved a disproportionately high classroom value, making this domain one of the clearest cases of low cultural distance.

\subsection{Mid Cultural Distance: Considerable Effort from Teachers to Use Current GenAI to Get Satisfactory Results}

Mid-cultural distance tasks arise when GenAI provides a useful starting point but fails to align cleanly with the demands of K--12 classrooms. Unlike low-distance tasks such as communication or activity brainstorming, here outputs could not be fixed with minor edits. Teachers expected ready-to-use, level-appropriate, or culturally resonant materials, but instead faced mismatched difficulty, formal vocabulary, or vague cultural references. To make the results usable, they had to invest in deliberate prompting, sustained revision, or supplement outputs with their own pedagogical and cultural expertise.

Two domains illustrate this: generating assessment questions (see M-1), where teachers adjusted difficulty and vocabulary through prompts, editing, or specialized tools; and designing culturally relevant learning activities (see M-2), where teachers reframed prompts or reworked outputs with local knowledge to compensate for underrepresented cultures. In both cases, GenAI was helpful but only after considerable effort, making these tasks clear examples of mid-cultural distance.

\subsubsection{When Teachers Generate Assessment Questions, Different GenAI Products Require Varying Levels of Prompting to Match the Desired Learners' Levels}

Assessment question generation was one of the most common GenAI use cases in K-12 teaching, mentioned by our participants. Teachers expected these tools to reduce the heavy workload of producing multiple question variants for different learners and levels of understanding. Many first learned about GenAI's potential in this domain through social media or colleagues' recommendations and hoped it would provide ready-to-use, curriculum-aligned questions.

In practice, however, experiences were mixed, revealing a noticeable cultural distance. Initial outputs often missed the mark, most commonly through mismatched difficulty levels or inaccessible language. For instance, U6, a U.S. high school teacher, found most outputs too simplistic: \textit{``Usually the ChatGPT is not going to do that. At least in my experience, it's just more going to be factual recall, which would be nice at the elementary level... but at the high school level, it's just not rigorous enough.''} By contrast, U5, a special education teacher, encountered overly advanced content for her third graders: \textit{``It's like algebra 3. Something crazy like solve for X, you know. And I'm like, oh yeah, that's not for you, honey. You don't need to learn that yet.''} Beyond difficulty, several teachers (e.g., S6, T4, U2) criticized the vocabulary used. As U2 noted: \textit{``Questions ChatGPT generated are generally pretty formal. And then it uses high-level vocabulary and high-level sentence structures that are typical of academic writing.''}

Teachers responded to these mismatches in three main ways. Some adjusted their prompts by inserting explicit keywords (e.g., ``college-level'') to push difficulty upward (U6), or inserted exact textbook content into the prompt (S10). Others manually revised the outputs to suit their students' needs. T6 explained: \textit{``I have to change the details all by my own, and just let ChatGPT do the formatting. Because ChatGPT can never know the nuances of each learner.''} A third group, often with institutional support, turned to specialized GenAI tools such as MagicSchool (U2, U8) or Mindjoy (S8), which were already fine-tuned to K--12 curricula and required far less iterative refinement.

Overall, teachers' experiences in assessment design illustrate a \textbf{mid cultural distance (M-1)} task. Their expectations of one-click, level-appropriate outputs were rarely met by general-purpose models, which defaulted to generic question types with mismatched difficulty and formal vocabulary. Yet unlike high-distance cases, these gaps could often be bridged through deliberate prompting, moderate editing, or the use of education-oriented GenAI tools. This made assessment generation both promising and demanding: the outputs saved time once refined, but they also required teachers' sustained effort to reach the desired pedagogical fit.

\subsubsection{When Teachers Generate Culturally Relevant Learning Activities, GenAI Requires Highly Effortful Prompting to Integrate Local Culture}

Culturally relevant learning activities in K-12 education are those that connect classroom content with students' everyday lives, traditions, and values. Teachers need to use them to make abstract concepts more relatable, to validate diverse identities, and to sustain student engagement. This work is especially important in multilingual or multicultural classrooms, where lessons that acknowledge local culture can reduce barriers to understanding and foster stronger connections between teachers and learners. Many teachers in our study hoped that GenAI could help them brainstorm such activities, expecting the systems to generate examples and scenarios that felt organically aligned with their students' lived experiences.

In practice, however, GenAI frequently fell short for underrepresented cultures in its training data. S2, a South African teacher, described the problem as \textit{``as far as cultural integration comes into AI, there's an element of vagueness.''} T2 in Taiwan, whose class included one-third Indigenous students, was more direct: \textit{``it is impossible for ChatGPT or Doubao (a popular GenAI application originated from Mainland China) to understand indigenous culture in Taiwan, because if something doesn't show up enough in the big databases, the AI can't grab or learn from it.''} Similarly, U8, a U.S. teacher, encountered imbalances when generating visuals: \textit{``especially when you start talking about children of color, you just don't have those options. You don't have those images generated.''}

To compensate, teachers invested significant effort in two main ways. First, some reframed their prompts to push GenAI toward more culturally relevant results. For example, U8 described how she deliberately referenced Girls Who Code when prompting ChatGPT, steering the model to highlight women in technology and counteract its default gender bias. Second, others worked directly on GenAI's outputs, revising and supplementing them with their own expertise. S2 explained: \textit{``AI can only say so much. But I know how I grew up, and how my students are growing up in their homes. So I bring that in myself to make it more relatable.''} In practice, she took GenAI's general suggestions and reworked them through a dramatic analogy she uses in class, her own situated framing of local understandings of marriage, to connect Shakespeare's themes to her students' lives: \textit{''I try to make it more relatable in the South African context, where in many cultures divorce is not allowed. Instead of divorce, women may opt to poison their husbands. . . in Zulu culture, once you are married, you cannot easily break the ties. So by drawing parallels to our own cultural practices, the text feels less distant and more understandable.''}

These cases show that generating culturally relevant learning activities is a \textbf{mid cultural distance (M-2)} task. Teachers' expectations of culturally attuned materials often went unmet, forcing them to expend considerable effort in reworking prompts, supplementing outputs with local knowledge, or strategically steering the model. While such strategies produced usable lessons, they also highlighted the limits of user effort when training data lacked sufficient representation. In the short term, teachers can act as cultural mediators; in the long term, reducing this cultural distance will require broader investments in more diverse datasets and infrastructures that better support underrepresented communities.

\subsection{High Cultural Distance: Poor Results Persist with Current GenAI Despite Teachers' Effort}

In some cases, teachers found that no matter how much effort they invested, whether by reframing prompts, editing outputs, or supplementing with their own expertise, GenAI still failed to produce usable results. We categorize these as instances of high cultural distance, where the gap between GenAI's default repertoire and classroom needs cannot be bridged by user adaptation alone. Two recurring sources drove such outcomes in our study: first, the extreme underrepresentation of certain languages and cultural traditions in training data (see H-1); and second, policy restrictions (see H-2) that blocked responses regardless of educational intent. Because our participants were frequent users who were disposed to keep trying, the high-distance cases we report are likely to understate how often teachers in general meet a gap they cannot close; a less persistent user would reach the point of abandonment sooner.

\subsubsection{When Local Languages and Traditions Are Missing from Training Data, GenAI Becomes Unusable}

Culturally relevant activities are central to teaching, but here the distinction from M-2 is crucial: while in mid-distance cases teachers could still bend GenAI outputs toward local relevance with sustained effort, in high-distance cases their efforts failed entirely. Teachers entered with the same expectation that GenAI could help them design materials rooted in students' cultural and linguistic realities. However, they quickly discovered that for extremely low-resource languages and cultures, the systems either produced incorrect responses or none at all. This marks the shift from mid cultural distance (where extra effort closes the gap) to high cultural distance (where the gap remains unbridgeable).

While identifying similarities between minority cultures and AI's default cultural assumptions, or explicitly prompting for characteristics of a specific ethnic group, can help bridge some of this cultural distance, the problem becomes significantly more challenging when language is involved. This is especially true when the local language needs to be used, or when aspects of the culture have not been widely verbalized in high-resource languages. For example, as an English and Sepedi (one of the South African official languages) teacher, S1 attempted to interact with ChatGPT in Sepedi rather than English, because English is not the native language for most students, and S1 wanted to use AI to explain concepts in the student's home language. After multiple trials with ChatGPT, however, he found the responses to be semantically and grammatically inaccurate, which discouraged him from using the local language again: \textit{``I also tried to use it in another official language in South Africa. But I realized that it's not really accurate... sometimes it gives me the answer in Sesotho. Sepedi and Sesotho are almost similar, but both are official languages. Still, the answers weren't accurate: when I read them, they didn't make sense. The language was wrong, or it was used in the wrong way. It was totally off, totally disappointing, so I didn't bother anymore.''}

S1's experience was not by chance. In fact, most South African teachers we interviewed had tried communicating with GenAI tools in local languages but were unable to obtain usable responses. As S6 noted: \textit{``When you send a question on Meta AI in Afrikaans (also a South African official language), it originally tried to generate it word by word, and at the end, it deleted everything and said, `We can't help you with this right now.'''}

These frustrations with local language support were echoed in other regions, where teachers confronted similar limitations around culturally specific knowledge. Taiwanese teacher T10 also described challenges when preparing materials about local Indigenous food and daily life for her young students: \textit{``Some groups eat a kind of raw meat marinated with millet called Tmmyan, De-Mo-Mian or} \begin{CJK}{UTF8}{bsmi}得麼面\end{CJK}\textit{. In my village, however, we eat a different version called Ge-Mo-Lian, or} \begin{CJK}{UTF8}{bsmi}哥麼鏈\end{CJK}\textit{, which is prepared in a similar way but with a shorter marination period.''} ChatGPT, however, was unable to recognize or articulate the nuanced differences between these two culturally specific foods.

The causes of this cultural distance are similar to those discussed in M-2, namely the lack of representation in training data. However, in this case, the challenge is not only more severe but also fundamentally different in kind. Whereas in mid-distance cases, teachers could still compensate by reframing prompts or supplementing outputs with their own cultural knowledge, here the scarcity of resources related to local languages and Indigenous traditions left them with no workable pathway to adaptation. As T10 explained, \textit{``There is little relevant information online; usually we consult the elders in the village to confirm culture and norms.''} Indeed, when our research team searched for the expressions she mentioned, we found almost no documentation available, either textual or audiovisual. We therefore classify these tasks as instances of high cultural distance: despite teachers' efforts, the gap could not be bridged without structural changes such as more inclusive training datasets. This left teachers not just frustrated but effectively powerless, underscoring their reliance on external actors (e.g., AI companies, policymakers) to address these fundamental representational gaps.

To illustrate how this powerlessness manifested in practice, South African teachers pointed out that the two most widely used GenAI applications, ChatGPT and Meta AI, handled low-resource language input in very different ways. ChatGPT produced responses riddled with errors, while Meta AI generated and then immediately retracted its replies with a refusal message (\textit{``Sorry, I'm currently not able to answer this question.''}). These differences likely stemmed from distinct QA and HCI design choices across companies, yet teachers' reactions were strikingly similar: \textit{``disappointment''} (S1, S6), \textit{``madness''} (S6), and, ultimately, \textit{a reversion to English} (S1, S6). If anything, ChatGPT's flawed outputs sometimes encouraged teachers to attempt a few more turns, whereas Meta AI's immediate refusal led them to abandon local language use altogether. T10 in Taiwan expressed similar frustration but also a desire for improvement. As she reflected: \textit{``I think it would be great if it could give me the correct information, because I'm not a tribal member myself. If the AI could provide accurate information, it would make preparing materials much easier, since I wouldn't always need to consult community members for verification. Of course, verification is important, but if the AI were accurate, it would greatly improve both the correctness and ease of lesson preparation for Indigenous students.''}

These cases from South Africa and Taiwan show \textbf{high cultural distance (H-1)}: when languages and cultures are extremely low-resource, the resulting gap is one that teachers cannot overcome on their own. Their experiences underscore the need for GenAI tools to take greater responsibility in representing marginalized languages and cultures, so that teachers can better support education in underrepresented communities and contribute to the preservation of cultural diversity.

\subsubsection{When Teachers' Queries Are Blocked by Policy-Level Controls, GenAI Becomes Unusable}

Teachers often approach GenAI tools with the expectation that they will provide open and reliable responses to support lesson preparation. They hope to gather examples, explanations, or culturally relevant references that can enrich their classrooms. Yet, in practice, these expectations are sometimes thwarted by policy-level restrictions. Unlike mid-distance cases, where additional effort or prompting might resolve misalignment, teachers here find themselves unable to obtain usable output because of policy-level restrictions, creating a form of high cultural distance.

Such restrictions arise at two different loci, but in both the obstacle reflects situated assumptions about what counts as appropriate classroom content, one built into the product, the other set by the institution that governs its use.

The clearest case lies in the policies embedded within GenAI products themselves. T2, a primary school teacher in Taiwan, used tools developed in mainland China, such as Doubao (``\begin{CJK}{UTF8}{bsmi}豆包\end{CJK}'') and ERNIE Bot (``\begin{CJK}{UTF8}{bsmi}文心一言\end{CJK}''), alongside ChatGPT. Because these products are trained largely on Chinese-language data, they can be an appealing option for Chinese-speaking teachers in Taiwan. Yet ERNIE Bot declined to answer some Taiwan-related queries, even though his classroom use was purely educational: \textit{``The language system, expressions, and even writing conventions all come from Mainland China. So, when we examine this issue, it is not really about grammar, but about where the products come from. Our use is constrained by the political tensions.''} Here the constraint sits inside the model's default repertoire: the product's content policies shape which Taiwan-related information it returns, and no amount of prompting or reframing on T2's side can change this.

A second, more expansive case locates the same kind of assumption outside the product, in the institution that surrounds it. U4, a high school humanities teacher from Texas, explained that his school had prohibited both teachers' use of AI and the teaching of AI to students. While no official rationale was provided, he tied it to the broader local politics around curriculum content: \textit{``Texas is an interesting state. We are the number one book banning district in the entire country. We banned Anne Frank. Anything about slavery, Frederick Douglass... there are definitely political undertones to it. I think it's kind of like this fear of having to control everything my students see. And teachers are gonna you know, there's this idea that teachers indoctrinate students.''} The mechanism here differs from T2's, since what is foreclosed is not a particular output but the tool as a whole, yet its source is continuous with it. GenAI's default repertoire is open-ended, spanning topics that local policy treats as contested. The distance in this case is therefore the mirror image of the language gaps in H-1: there, the model's defaults sat too far from local demands because they \textit{omitted} local knowledge; here, the model's defaults are foreclosed because their openness extends beyond what local policy permits in the classroom. In both, the gap runs in the same direction, from a globally trained system's defaults toward a locally situated demand, and in both it lies beyond what any individual teacher can close.

Policy-level controls are cultural not because all policy is culture, but because policies reflect assumptions about appropriate topics and classroom uses. In T2's case, the problem was not simply that a query was blocked, but that the product's content policies shaped which Taiwan-related information was available for classroom use. Since teachers cannot control either local educational regulations or the policy filters built into GenAI products, these restrictions represent a form of high cultural distance. As T2 observed, the source of the product itself matters: teachers must recognize where a GenAI system was created in order to anticipate its blind spots or restrictions. Even when switching to locally developed applications seems to reduce the gap, those systems still carry the biases and filters of their own contexts. For teachers, policy-level restrictions are not gaps they can close through prompting or adaptation, but structural barriers beyond their control. Even when teachers recognize and critique these constraints, their efforts rarely lead to change, leaving them bound by policies they cannot influence.

In sum, policy-level controls illustrate a kind of \textbf{high cultural distance (H-2)} that teachers cannot bridge through their own effort. Whether imposed by schools or encoded into products, such restrictions effectively block teachers' attempts to adapt GenAI, underscoring how policy and institutional settings shape the boundaries of classroom AI use.

\section{Discussion}

Our study examined how K--12 teachers in South Africa, Taiwan, and the United States engaged with GenAI in their everyday classroom practice. While prior work has focused on adoption rates, infrastructure gaps, or ethical concerns, our analysis introduces the concept of \textit{cultural distance} as a way to capture the varying degrees of fit between GenAI's global defaults and teachers' local classroom demands. As our findings show, cultural distance provides a spectrum-based vocabulary for analyzing not just whether GenAI is aligned, but how much effort users must expend to make it usable, and where those efforts succeed, fail, or are blocked by structural barriers.

In the following subsections, we first draw out four properties of cultural distance that the findings support and that distinguish it from existing ways of talking about bias, localization, or alignment. We then state how the framework can be used: the conditions under which it applies, the steps for applying it, the circumstances in which it should be adapted or extended, and the cases in which it should not be used. Next, we state the concept's two theoretical contributions: to AI alignment research, by foregrounding cultural fit as a dimension of alignment, and to HCI research, by making the labor of adaptation comparable across tasks. Finally, we draw out its practical contribution in the form of design and policy implications, highlighting how cultural distance can guide both system development and institutional governance toward more inclusive and context-sensitive AI practices.

\subsection{Four Properties of Cultural Distance Drawn From the Findings}

We defined cultural distance as the gap between GenAI's default cultural repertoire, meaning the globally available training data, embedded norms, and product-level policies built into these systems, and the situated demands of teachers' everyday practice. The findings give that definition empirical content. Here we draw out four properties of the concept that the findings support, each anchored to the cases that show it, and that distinguish cultural distance from existing ways of talking about bias, localization, or alignment.

First, cultural distance is task-specific, not a property of a tool or a culture. The same teacher using the same model can sit at very different distances depending on the task: near in making slides more colorful for students, far in representing an Indigenous food tradition. Distance is therefore not a fixed attribute of ``ChatGPT'' or of ``Taiwanese classrooms,'' but a relation that is re-established task by task. This is what lets the concept discriminate within a single teacher's practice, where coarser notions of national or model bias cannot. Table 1 shows it directly: the same teacher (for example, S8, T10, or U6) appears at two or three different levels depending on the task.

Second, cultural distance is directional. Unlike the symmetric, nation-to-nation construct in international business, our distance runs specifically from a globally trained artifact's defaults \textit{toward} a local classroom's demands. The asymmetry matters analytically: it locates the burden of adaptation on the teacher's side and the source of the gap on the system's side, making visible who is doing the work of closing it. The high-distance cases in H-1 and H-2 show this most clearly: in each, the burden fell on the teacher, while the system's defaults, or the rules governing its use, did not move.

Third, cultural distance is enacted through labor. Distance becomes observable through what teachers \textit{do}: prompting, editing, supplementing with their own cultural knowledge, switching tools, or, at the limit, abandoning the task. The magnitude of distance is legible in the kind and amount of this labor, which is why the concept foregrounds adaptation work that bias-or-localization framings tend to leave invisible. The low, mid, and high subsections of the Findings are organized precisely by this labor, from minor edits (L-1, L-2) through sustained prompting and revision (M-1, M-2) to abandonment (H-1, H-2).

Fourth, cultural distance is structurally bounded, and can be estimated before the fact. Some gaps lie beyond what any teacher effort can close, and resolving them requires change at the level of datasets, policy, and infrastructure rather than user skill. Crucially, although distance is \textit{enacted} through labor, it need not be diagnosed only \textit{retrospectively}, by waiting to see whether effort eventually succeeds. A task's likely distance can be estimated in advance from properties independent of any one teacher's outcome: whether it depends on a low-resource language, on knowledge that remains largely undigitized, or on content subject to hard policy filters. These structural markers let researchers and designers anticipate high-distance cases before teachers have exhausted their effort, and they keep the concept from collapsing into the tautology that ``high distance is whatever effort fails to fix.'' The two high-distance forms illustrate the two kinds of marker: unsupported languages and undigitized knowledge (H-1) and policy filters (H-2) were identifiable before any particular prompt was tried.

Together, these properties offer an explanation for why the spectrum recurred across the three contexts in our sample despite their very different infrastructures, languages, and policy regimes: on this account, cultural distance is not tied to any one region or school system but to the structural relationship between globally trained systems and locally situated practice. Whether the spectrum recurs in other contexts remains to be examined. On this account, the concept also reframes teachers' role. Teachers appear as the actors on whom the labor of alignment unevenly falls: sometimes lightly, sometimes heavily, and sometimes beyond what any effort can carry. Cultural distance thus does not only describe a problem; it makes visible how the work of alignment is distributed between global systems and local users, and clarifies where responsibility for closing the gap must ultimately lie.

\subsection{Using the Cultural Distance Framework: When It Applies, How to Apply It, and When to Adapt or Set It Aside}

Cultural distance is meant to be used by future researchers studying AI in situated practice, and this section states how. We first state the conditions under which the framework is best suited and the steps for applying it. We then describe settings that meet the conditions above but differ enough from ours to require extending or reworking parts of the framework, and finally the questions it is not designed to answer. Figure 3 summarizes these conditions, steps, and limits.

\begin{figure}[htbp]
\centering
\includegraphics[width=\textwidth]{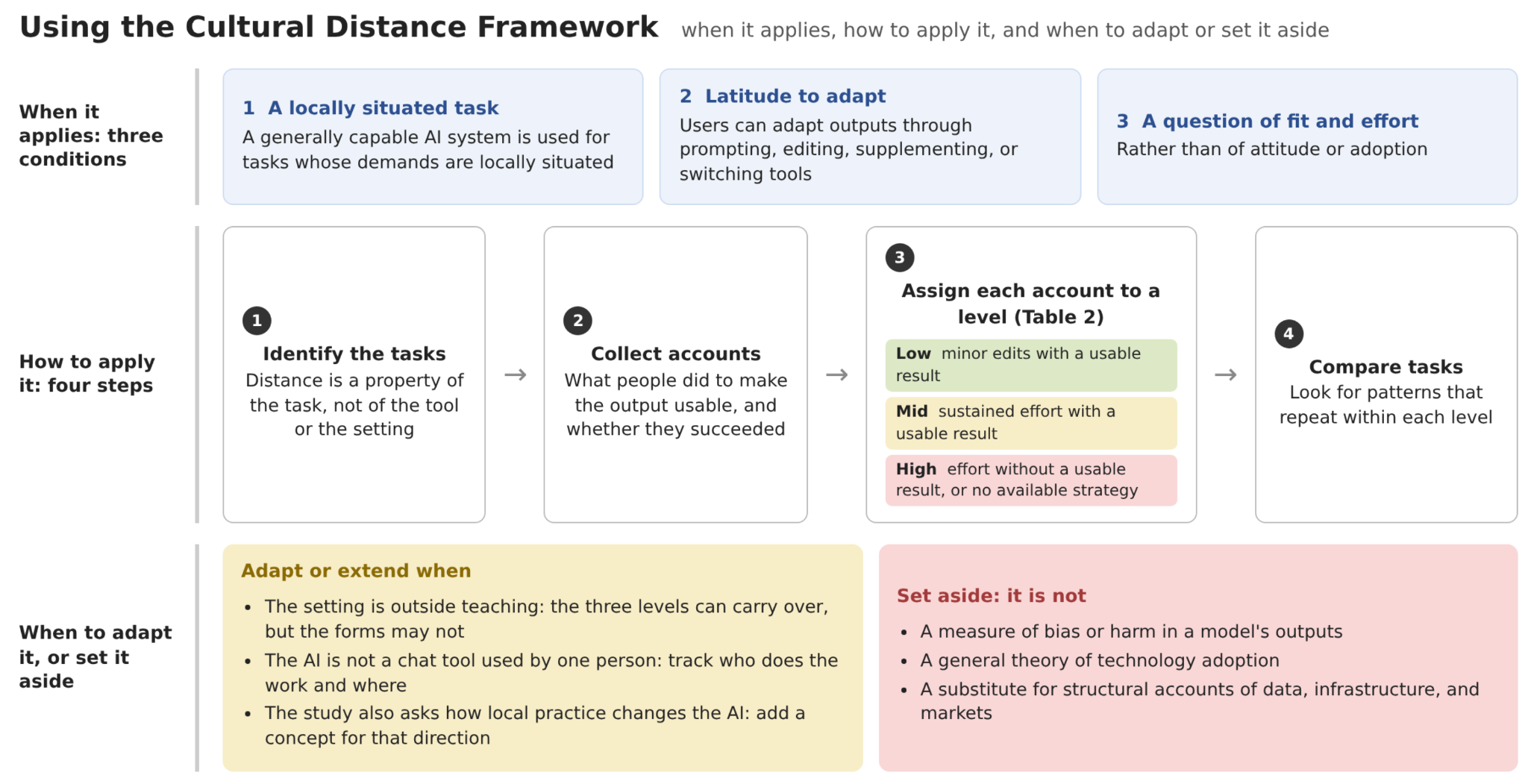}
\caption{How to Apply the Cultural Distance Framework}
\label{fig:3}
\end{figure}

The framework is best suited to settings where three conditions hold. First, a generally capable AI system is being used for tasks whose demands are locally situated, so that there is a default repertoire on one side and a situated demand on the other. Second, the people using the system have some latitude to adapt its outputs, through prompting, editing, supplementing, or switching tools, so that effort is a meaningful variable; where outputs are consumed as given, with no adaptation possible, there is nothing for the framework to measure. Third, the question of interest is one of fit and effort rather than of attitude or adoption. It is not about whether people take up the tool or how they feel about it, but how far its outputs sit from what they need and what it costs them to close that gap. K-12 teaching met all three conditions, but so do many other settings in which professionals use GenAI for situated work, such as journalists writing for a local readership, clinicians documenting care in a regional system, or civil servants drafting in a minority language.

When these conditions hold, applying the framework involves four steps, which follow the way we developed it. The first is to identify the tasks for which the AI system is used, since distance is a property of the task and not of the tool or the setting. The second is to collect accounts of what people did to make the output usable for each task and whether they succeeded, since distance is enacted through labor and becomes visible only through it. The third is to assign each account to a level using the criteria in Table 2: minor edits with a usable result (low), sustained effort with a usable result (mid), or effort without a usable result or no available strategy (high). The fourth is to compare tasks and look for patterns that repeat within each level. It is also worth noting what makes a task likely to fail from the start, such as needing a language the AI does not support, knowledge that is not online, or a topic the AI refuses to discuss, so that such tasks can be spotted before anyone spends effort on them. Tables 2 and 3 and Figure 2 show what the output of these steps looks like for our case.

Some settings meet the three conditions but differ enough from ours to require extending or reworking parts of the framework. This happens in three cases: when the setting is outside teaching, when the AI is not a chat tool used by one person, and when the study also asks how local practice changes the AI. First, the six forms we report come from teachers in our sample, so a study in another field should expect different forms; the three levels can carry over, but the forms may not. Second, the criteria in Table 2 assume one person adapting a chat tool's output. If the AI is built into an institution's workflow, or if a group adapts its output together, the criteria will need to track who does the work and where. Third, our framework only looks at how far the AI's defaults sit from local needs. A study that also wants to know how local use changes what the AI treats as default will need to add a concept that captures that direction.

Finally, even in settings that meet the three conditions, the framework should not be used to answer questions it was not built for. There are three such questions. It is not a measure of bias or harm in a model's outputs; a model can be culturally distant from a classroom without producing anything biased, and biased output that a teacher can easily edit out is low distance, not high. It is not a general theory of technology adoption, and it has nothing to say about why a teacher takes up GenAI in the first place or stops using it for reasons unrelated to fit. And it is not a substitute for the structural accounts of data, infrastructure, and markets on which it draws: those accounts explain why defaults sit where they do, whereas cultural distance describes what that placement costs the user. Used within these limits, the framework gives a way of asking a question that the existing accounts leave open, which is how much of the work of alignment falls on the user and whether that work can succeed.

\subsection{Theoretical Contributions of Cultural Distance}

The concept makes two theoretical contributions, one to AI alignment research and one to HCI research on invisible labor.

\subsubsection{Extending AI Alignment through Cultural Distance}

Our concept of cultural distance also extends ongoing debates about AI alignment. Alignment research has largely concentrated on the system side, i.e., fine-tuning models to reflect human values around ethics, safety, or governance (Gabriel, 2020; Ge et al., 2024; Hadfield-Menell \& Hadfield, 2019), with more recent work shifting attention to the user side, where alignment is continually worked out through correction, rejection, and mutual adaptation in everyday use (Fan et al., 2025; Shen, Knearem, et al., 2024; Shen, Ma, et al., 2024). Cultural distance builds on this user-side turn but presses it in a direction neither strand has addressed: the question of \textit{cultural} fit.

It shows that alignment is not only about universal values but also about how well AI systems fit into the everyday cultural contexts where they are used. In our study, teachers across South Africa, Taiwan, and the United States demonstrated that GenAI could be perfectly safe, fluent, and coherent yet still misaligned with classroom needs. Sometimes it generated questions at the wrong difficulty level, overlooked Indigenous traditions, or declined certain queries under content policies. These cases illustrate that value alignment cannot be separated from cultural alignment: a system is not ``aligned'' if users must continually overcompensate for its gaps, or if some tasks lie structurally beyond reach no matter how much effort they invest.

This is where cultural distance makes a distinctive middle-range contribution to alignment debates. Abstract principles like fairness, safety, and accountability specify what alignment should ultimately secure but say little about how it is achieved task by task; individual case studies show alignment breaking down in one setting, but do not travel to others. Cultural distance sits between them, giving alignment research a vocabulary for \textit{how much} and \textit{what kind} of fit a given task demands, and thus for comparing alignment work across tasks, tools, and regions.

Because cultural distance distinguishes gaps teachers can close from those they cannot, it also relocates responsibility. Where adaptation is possible, alignment is partly a matter of user practice; but where it is structurally foreclosed, as with unsupported languages, undigitized knowledge, and hard policy filters, the obligation shifts from the individual teacher to those who shape the defaults: AI developers who set training data and product policy, and policymakers who govern access and infrastructure. In this sense, cultural distance shows why alignment cannot be solved at the model level alone, and why user adaptation must itself be treated as part of the alignment process rather than as a sign that alignment has already been achieved.

This framing also opens concrete directions for future research. The concept can be used to \textit{compare across settings}, mapping how much adaptive work different users (e.g., teachers, journalists, clinicians) must perform to make the same systems usable, and to \textit{diagnose failure points}, distinguishing misalignments that users can resolve through prompting and editing (mid-distance) from those that require structural change (high-distance). More broadly, it invites researchers to test whether the low-to-high spectrum we observed in education recurs in other professional domains where globally trained systems meet locally situated practice.

\subsubsection{Contributions to HCI: From Visible to Comparable Labor}

For HCI, the concept of cultural distance offers a new way to understand how global technologies intersect with local practices. Prior work has examined technology adoption in schools, workplaces, and communities, often focusing on usability, infrastructure, or individual attitudes toward innovation (Behrend et al., 2011; FakhrHosseini et al., 2024; Straub, 2009; Q. Xiao et al., 2025). Our study shifts attention to the cultural dimension: the degree of effort required for users to make global systems meaningful in their situated contexts. We see two main contributions.

First, cultural distance contributes a comparative vocabulary for analyzing cross-cultural technology use. It highlights how technologies may align more easily with certain practices (low distance), require sustained adaptation (mid distance), or remain structurally out of reach (high distance). For example, teachers across South Africa, Taiwan, and the United States experienced similar ease in communication tasks but diverged sharply when local languages or Indigenous traditions were involved. This comparative lens helps HCI researchers identify not only whether a system works but also how much work different users must do to make it work.

Second, cultural distance does more than add another case of invisible labor to a literature that has long made such labor visible (Calacci \& Pentland, 2022; Dangol et al., 2026; Harmon \& Silberman, 2019; Q. Xiao, Fan, et al., 2026; Q. Xiao, Hu, et al., 2026). That teachers expend hidden effort to adapt to GenAI, by editing assessments, reframing prompts, and supplementing with cultural expertise, is, by now, an expected finding rather than a novel one. Our contribution is to render this labor comparable: cultural distance turns adaptation effort into a graded quantity that can be located on a spectrum, read off the kind and amount of work a task requires, and set side by side across tasks, tools, and regions. Where prior work makes invisible labor \textit{seen}, cultural distance makes it \textit{comparable} (this task is farther than that one; this gap closes with effort while that one does not).

This comparative move also distinguishes our lens from recent studies of teachers' situated labor around classroom AI. Most directly, Yin et al. (2026) characterize teachers as everyday \textit{repairers} who restore Responsible AI values (fairness, accountability, relational care) when classroom AI systems break down, drawing on repair and maintenance theory to make this labor visible. Our lens is complementary but differs in object, granularity, and scope. Where pedagogical repair centers on value breakdowns and the work of mending them, cultural distance asks a prior and more graduated question: how far a system's default outputs sit from local classroom demands in the first place, and how much effort, from minor edits to gaps that cannot be bridged at all, their alignment requires. Repair describes what teachers do once a system has failed or presented conflicting values; cultural distance measures the spectrum of fit that determines whether, and how hard, such repair must even be attempted. Their account also foregrounds Responsible AI value tensions within a single national setting, whereas ours foregrounds cultural and linguistic fit across three regions.

More fundamentally, the two framings differ in a deeper assumption. Frameworks built around repair or value alignment take \textit{alignment as the goal}, and that starting point quietly narrows what they can see. If the aim is always to bring a system back into accord with teachers' needs, then the only situations that come into focus are those where such accord is both necessary and achievable, where a gap exists. What drops out of view are the two ends of the spectrum: cases where no alignment work is needed because the system's defaults already fit, and cases where no amount of work will close the gap because it is structurally foreclosed. Cultural distance, by gauging fit, holds all three in a single frame: the low-distance tasks where alignment is unnecessary, the mid-distance tasks where it is effortful but reachable, and the high-distance tasks where it is out of reach no matter the effort. It further opens, as a conceptual matter, the question of whether teachers might in some cases reasonably decline to adapt the system at all, setting the tool aside to preserve a local practice rather than expending further effort to bend GenAI's outputs toward it, a possibility an alignment-as-goal framing is not equipped to name, and that future work could examine directly. By treating distance as the prior question, cultural distance extends HCI's analytic toolkit not only for analyzing how teachers close the gap, but for recognizing when the gap need not, or cannot, be closed at all.

\subsection{Practical Contribution: Design and Policy Implications}

Our framework of cultural distance also points to concrete directions for design and policy. Because cultural distance captures both the gap between GenAI defaults and local practices, and the effort required for teachers to bridge that gap, it provides a diagnostic lens for identifying where intervention is most needed and what form it should take. Table 4 summarizes four concrete mechanisms, each anchored to the findings that motivate it.

\begin{table}[htbp]
\centering
\small
\setstretch{1.0}
\caption{Four Design and Policy Mechanisms Derived From the Cultural Distance Framework, Each Anchored to the Empirical Findings That Motivate It}
\label{tab:4}
\begin{tabularx}{\textwidth}{@{}>{\raggedright\arraybackslash}p{3.1cm} >{\raggedright\arraybackslash}p{1.6cm} >{\raggedright\arraybackslash}X >{\raggedright\arraybackslash}p{2.4cm}@{}}
\toprule
\textbf{\textbf{Mechanism}} & \textbf{\textbf{Distance level}} & \textbf{\textbf{Motivating finding}} & \textbf{\textbf{Primary responsibility}} \\
\midrule
Teacher-facing localization controls & Mid & Teachers re-improvised grade level, difficulty, and cultural framing through repeated prompting and manual editing (M-1, M-2). & Developers \\
\addlinespace
Community verification workflow & High & Tools hallucinated or could not represent undigitized local/Indigenous knowledge; teachers fell back on community elders (H-1). & Developers \\
\addlinespace
Policy transparency \& appeal mechanism & High & Educational queries were blocked without explanation or recourse, by product filters or institutional bans (H-2). & Developers \& policymakers \\
\addlinespace
Cultural distance audit & All levels & The recurring low--mid--high spectrum supplies the dimensions (language, curriculum, local knowledge, policy sensitivity) a pre-deployment audit should probe. & Developers \& policymakers \\
\bottomrule
\end{tabularx}
\end{table}

\subsubsection{Design Implications}

At the low cultural distance level, teachers benefited from GenAI's fluency and flexibility with minimal adaptation. Here, the design challenge is less about new features and more about sustaining reliability: ensuring that tools remain accessible, easy to integrate, and responsive to the everyday pace of teaching. At the mid level, teachers' heavy prompting and editing point to a specific opportunity. Systems could offer teacher-facing localization controls: first-class settings for grade level, curriculum standard, language variety, and community context that persist across a teacher's sessions, so that adjustments teachers now re-improvise through prompts each time, e.g., inserting ``college-level'' to raise difficulty (M-1) or steering toward references like Girls Who Code to counter default bias (M-2), are set once. At the high level, where teachers' efforts could not bridge the gap, the priority shifts from generating output to representing limits honestly. When a query depends on undigitized local or Indigenous knowledge, a \textit{community verification workflow} that flags the answer as ``uncertain---consult a local expert'' would serve teachers far better than a confident hallucination, as when ChatGPT could not distinguish the two Indigenous food traditions our participant described (H-1). Across all levels, GenAI design should foreground teachers' time and cognitive load, making adaptation less labor-intensive and more collaborative.

\subsubsection{Policy Implications}

Cultural distance also underscores that alignment cannot be left to teachers alone. At the structural level, policies play a decisive role in widening or narrowing the gap. Investment in local AI research, language resources, and educational data infrastructures can reduce high cultural distance by giving teachers tools that better reflect their contexts. Equally important is governance of how systems and institutions handle restrictions. Teachers in our study who met such restrictions received no explanation, whether from product-level filters (e.g., ERNIE Bot declining Taiwan-related queries) or institutional bans (H-2), leaving them unable to adapt or critique effectively. We therefore call for \textit{policy transparency and appeal mechanisms}: when an educational query is blocked, systems and institutions should communicate why, offer an alternative path where possible, and provide an educator review channel. More broadly, a \textit{cultural distance audit}, probing a tool across languages, curriculum standards, local knowledge, and policy-sensitive content before deployment, would let developers and procurement bodies anticipate high-distance failures in advance.

\subsubsection{For Both Developers and Policymakers}

Ultimately, cultural distance highlights the uneven distribution of alignment labor. Where gaps are small, teachers can adapt outputs with little effort; where they are large, adaptation is either burdensome or structurally foreclosed. Design and policy should therefore aim not only to improve models, but also to redistribute responsibility more fairly. This can be achieved by equipping teachers with supportive tools, by ensuring companies take responsibility for data diversity and policy transparency, and by enabling institutions to provide educators with training, infrastructure, and protections.

\section{Limitations and Future Work}

While our study advances the concept of cultural distance and grounds it in cross-regional teacher interviews, several limitations should be acknowledged. First, our empirical basis is modest in scale: 30 interviews, with 10 participants each from South Africa, Taiwan, and the United States. As we noted in the Context and Findings sections, this sample allowed us to capture contrasting perspectives across regions, but it cannot represent the variation within any one of them, let alone K--12 education more broadly. Our inclusion criteria, moreover, favored frequent GenAI users with an interest in cultural adaptation, which, as discussed in the High Cultural Distance subsection, may bias the sample toward teachers already disposed to adapt rather than abandon these tools, and thus toward the lower and middle reaches of the spectrum. Future work should extend this framework to additional cultural contexts, languages, and school systems, particularly regions with different infrastructural conditions or stronger domestic AI ecosystems, and to teachers less inclined to persist with GenAI.

Second, our analysis focused primarily on teachers' engagements with chat-based GenAI tools. While this reflects the most common form of GenAI adoption in classrooms today, it does not capture experiences with API integrations, multimodal tools, or specialized learning management systems. Future studies could explore how cultural distance manifests differently when GenAI is embedded deeper into institutional infrastructures or used alongside other emerging technologies.

Third, the data are self-reported through interviews rather than complemented with ethnographic observation or usage logs. While interviews provided rich accounts of expectations, practices, and challenges, future research could benefit from triangulation with classroom observations or longitudinal usage data to better capture how cultural distance evolves over time.

A further conceptual caveat concerns how our levels are bounded. Because cultural distance is enacted through teachers' effort, our coding of the mid/high boundary is in part outcome-informed; we partly recognized a case as high-distance through whether sustained effort ultimately failed. As we argue in the ``Four Properties of Cultural Distance'' subsection, distance can also be estimated prospectively from properties independent of any single outcome, which keeps the concept from reducing to its results; nonetheless, our analysis leans on teachers' retrospective accounts, and borderline cases may have been sorted partly with the benefit of hindsight. Triangulating future coding with prospective structural markers, fixed in advance of outcome, would tighten this boundary.

Finally, we present cultural distance as an exploratory framework rather than a closed typology. The three levels we describe (low, mid, and high) emerged from our empirical cases, along with the specific categories within those various degrees, but other forms may exist in different domains or with different user groups. Future work could test, refine, or extend this framework in other sectors such as journalism, healthcare, or creative industries, as well as examine how cultural distance interacts with factors like professional identity, institutional support, or national policy. A complementary line of work could take the framing of cultural or linguistic proximity, asking how close a given GenAI system already sits to a particular language or curriculum before any teacher effort is spent, and relating that starting point to the levels of distance we describe here.

In this sense, we encourage future researchers to take our framework not as a final model but as an invitation: to adapt, refine, and extend cultural distance in new domains, to test its relevance in other professions, and to use it as a springboard for theorizing how global AI systems intersect with diverse local practices. We hope this concept serves less as a conclusion than as a generative prompt for continued exploration.

\section{Conclusion}

This paper introduced \textit{cultural distance}, defined as the gap between GenAI's default cultural repertoire and the situated demands of K--12 teaching, as a framework for understanding why and how GenAI aligns or fails to align with classroom practice. Drawing on 30 interviews with teachers in South Africa, Taiwan, and the United States, we showed that cultural distance exists on a spectrum: low when outputs require only minor edits, mid when sustained prompting or revision is needed, and high when the gap is structural, rooted in unsupported languages, undigitized knowledge, or policy filters, and lies beyond what teacher effort alone can close. By distinguishing these levels, cultural distance provides a vocabulary for analyzing the uneven labor of alignment and extends current debates beyond system design to everyday practice. The framework is grounded in an exploratory study of 30 frequent GenAI users in three contexts, and we present it as a promising lens, inviting future research to examine it with larger and more diverse samples and in other domains where global AI systems encounter local practices.

\section*{Author Contributions Statement}

Qing Xiao, Ruiwei Xiao, Hanqi Jane Li, Phenyo Phemelo Moletsane, Xinying Hou, John Stamper, and Hong Shen contributed to the study's conception and design. Qing Xiao and Ruiwei Xiao took primary responsibility for study design, data collection, analysis, drafting and refining the manuscript. Xinying Hou, John Stamper, and Hong Shen provided guidance on the study's theoretical framing. Hanqi Jane Li and Phenyo Phemelo Moletsane contributed to the data collection and analysis. All authors reviewed and approved the final manuscript and take responsibility for the integrity and accuracy of the work.

\section*{References}
\begin{hangparas}{1.5em}{1}
\small

Ahn, D., \& Lim, H. (2025). Exploring K-12 physical education teachers' perspectives on opportunities and challenges of AI integration through ideation workshops. In \textit{Proceedings of the 2025 CHI Conference on Human Factors in Computing Systems} (pp. 1--16). ACM. \url{https://doi.org/10.1145/3706598.3713646}

Al Jazeera. (2024, September 18). \textit{What's South Africa's new school language law and why is it controversial?} \url{https://www.aljazeera.com/news/2024/9/18/whats-south-africas-new-school-language-law-and-why-is-it-controversial}

Alexander, M. (2025). \textit{The languages of South Africa}. South Africa Gateway. \url{https://southafrica-info.com/arts-culture/the-languages-of-south-africa/}

Awad, G. H., Patall, E. A., Rackley, K. R., \& Reilly, E. D. (2016). Recommendations for culturally sensitive research methods. \textit{Journal of Educational and Psychological Consultation}, \textit{26}(3), 283--303. \url{https://doi.org/10.1080/10474412.2015.1046600}

Behrend, T. S., Wiebe, E. N., London, J. E., \& Johnson, E. C. (2011). Cloud computing adoption and usage in community colleges. \textit{Behaviour \& Information Technology}, \textit{30}(2), 231--240. \url{https://doi.org/10.1080/0144929X.2010.489118}

Bender, E. M., Gebru, T., McMillan-Major, A., \& Shmitchell, S. (2021). On the dangers of stochastic parrots: Can language models be too big? In \textit{Proceedings of the 2021 ACM Conference on Fairness, Accountability, and Transparency} (pp. 610--623). ACM. \url{https://doi.org/10.1145/3442188.3445922}

Bernstein, A. (2025). \textit{Artificial intelligence (AI) in K-8 education: Understanding teachers' perceptions and district leader readiness while preparing for AI adoption} [Doctoral dissertation, Concordia University Chicago].

Blasi, D., Anastasopoulos, A., \& Neubig, G. (2022). Systematic inequalities in language technology performance across the world's languages. In \textit{Proceedings of the 60th Annual Meeting of the Association for Computational Linguistics (Volume 1: Long Papers)} (pp. 5486--5505). Association for Computational Linguistics. \url{https://doi.org/10.18653/v1/2022.acl-long.376}

Braun, V., \& Clarke, V. (2019). Reflecting on reflexive thematic analysis. \textit{Qualitative Research in Sport, Exercise and Health}, \textit{11}(4), 589--597. \url{https://doi.org/10.1080/2159676X.2019.1628806}

BusinessTech. (2024, August 26). \textit{South Africa ranks as one of the biggest users of ChatGPT and AI globally}. \url{https://businesstech.co.za/news/technology/787980/south-africa-ranks-as-one-of-the-biggest-users-of-chatgpt-and-ai-globally/}

Calacci, D., \& Pentland, A. (2022). Bargaining with the black-box: Designing and deploying worker-centric tools to audit algorithmic management. \textit{Proceedings of the ACM on Human-Computer Interaction}, \textit{6}(CSCW2), 1--24. \url{https://doi.org/10.1145/3570601}

Chitiga, M., Owusu-Sekyere, E., \& Tsoanamatsie, N. (2014). Income inequality and limitations of the Gini index: The case of South Africa. \textit{HSRC Review}, \textit{12}(5), 9--11. \url{http://hdl.handle.net/20.500.11910/2201}

Chounta, I.-A., Bardone, E., Raudsep, A., \& Pedaste, M. (2022). Exploring teachers' perceptions of artificial intelligence as a tool to support their practice in Estonian K-12 education. \textit{International Journal of Artificial Intelligence in Education}, \textit{32}(3), 725--755. \url{https://doi.org/10.1007/s40593-021-00243-5}

Cohn, C., Hutchins, N., Le, T., \& Biswas, G. (2024). A chain-of-thought prompting approach with LLMs for evaluating students' formative assessment responses in science. \textit{Proceedings of the AAAI Conference on Artificial Intelligence}, \textit{38}(21), 23182--23190. \url{https://doi.org/10.1609/aaai.v38i21.30364}

Cordero, J., \& Cordero-Castillo, A. (2025). Exploring the potential of generative AI in education: Opportunities, challenges, and best practices for classroom integration. In \textit{Foundations of computer science and frontiers in education: Computer science and computer engineering} (pp. 252--265). Springer. \url{https://doi.org/10.1007/978-3-031-85930-4_23}

Council of Indigenous Peoples. (2024). \textit{Population and demographics of Indigenous peoples in Taiwan}. \url{https://www.apc.gov.tw/portal/docList.html?CID=DC2E742A2C9F6E3A}

Cukurova, M., Miao, X., \& Brooker, R. (2023). Adoption of artificial intelligence in schools: Unveiling factors influencing teachers' engagement. In \textit{Artificial intelligence in education} (pp. 151--163). Springer. \url{https://doi.org/10.1007/978-3-031-36272-9_13}

Dangol, A., Kotiyal, S., Wolfe, R., Bowers, A. J., Vigil, A., Yip, J., Kientz, J. A., Shahid, S., Yeh, T., Cho, V., \& Davis, K. (2026). Relief or displacement? How teachers are negotiating generative AI's role in their professional practice. In \textit{Proceedings of the 2026 CHI Conference on Human Factors in Computing Systems} (pp. 1--21). ACM. \url{https://doi.org/10.1145/3772318.3791904}

Elnaem, M. H., Okuyan, B., Mubarak, N., Thabit, A. K., AbouKhatwa, M. M., Ramatillah, D. L., Isah, A., Al-Jumaili, A. A., \& Mohamed Nazar, N. I. (2025). Students' acceptance and use of generative AI in pharmacy education: International cross-sectional survey based on the extended unified theory of acceptance and use of technology. \textit{International Journal of Clinical Pharmacy}, \textit{47}(4), 1097--1108. \url{https://doi.org/10.1007/s11096-025-01936-w}

Ethnologue. (2025). \textit{South African Sign Language}. \url{https://www.ethnologue.com/language/sfs/}

Faisal, F., Wang, Y., \& Anastasopoulos, A. (2022). Dataset geography: Mapping language data to language users. In \textit{Proceedings of the 60th Annual Meeting of the Association for Computational Linguistics (Volume 1: Long Papers)} (pp. 3381--3411). Association for Computational Linguistics. \url{https://doi.org/10.18653/v1/2022.acl-long.239}

FakhrHosseini, S., Chan, K., Lee, C., Jeon, M., Son, H., Rudnik, J., \& Coughlin, J. (2024). User adoption of intelligent environments: A review of technology adoption models, challenges, and prospects. \textit{International Journal of Human--Computer Interaction}, \textit{40}(4), 986--998. \url{https://doi.org/10.1080/10447318.2022.2118851}

Fan, X., Xiao, Q., Zhou, X., Pei, J., Sap, M., Lu, Z., \& Shen, H. (2025). User-driven value alignment: Understanding users' perceptions and strategies for addressing biased and discriminatory statements in AI companions. In \textit{Proceedings of the 2025 CHI Conference on Human Factors in Computing Systems} (pp. 1--19). ACM. \url{https://doi.org/10.1145/3706598.3713477}

Filiz, O., Kaya, M. H., \& Adiguzel, T. (2025). Teachers and AI: Understanding the factors influencing AI integration in K-12 education. \textit{Education and Information Technologies}, \textit{30}(13), 17931--17967. \url{https://doi.org/10.1007/s10639-025-13463-2}

Gabriel, I. (2020). Artificial intelligence, values, and alignment. \textit{Minds and Machines}, \textit{30}(3), 411--437. \url{https://doi.org/10.1007/s11023-020-09539-2}

Gallup. (2025). \textit{Teaching for tomorrow: Unlocking six weeks a year with AI}. Walton Family Foundation. \url{https://static.waltonfamilyfoundation.org/df/fb/eba12807470a9402d7433cc47dba/teaching-for-tomorrow-unlocking-six-weeks-a-year-with-ai-report.pdf}

Gay, G. (2018). \textit{Culturally responsive teaching: Theory, research, and practice} (3rd ed.). Teachers College Press.

Ge, L., Halpern, D., Micha, E., Procaccia, A. D., Shapira, I., Vorobeychik, Y., \& Wu, J. (2024). Axioms for AI alignment from human feedback. \textit{Advances in Neural Information Processing Systems}, \textit{37}, 80439--80465. \url{https://doi.org/10.52202/079017-2557}

Google DeepMind. (2023). \textit{Gemini}. \url{https://deepmind.google/models/gemini/}

Grand View Research. (2025). \textit{AI in education market size, share \& trends analysis report by component, by deployment, by technology (NLP, ML), by application (intelligent tutoring system, learning platform \& virtual facilitators), by end-use, by region, and segment forecasts, 2025--2030}. Retrieved May 6, 2025, from \url{https://www.grandviewresearch.com/industry-analysis/artificial-intelligence-ai-education-market-report}

Hadfield-Menell, D., \& Hadfield, G. K. (2019). Incomplete contracting and AI alignment. In \textit{Proceedings of the 2019 AAAI/ACM Conference on AI, Ethics, and Society} (pp. 417--422). ACM. \url{https://doi.org/10.1145/3306618.3314250}

Hao, K. (2022, April 19). Artificial intelligence is creating a new colonial world order. \textit{MIT Technology Review}. \url{https://www.technologyreview.com/2022/04/19/1049592/artificial-intelligence-colonialism/}

Harmon, E., \& Silberman, M. S. (2019). Rating working conditions on digital labor platforms. \textit{Computer Supported Cooperative Work (CSCW)}, \textit{28}(5), 911--960. \url{https://doi.org/10.1007/s10606-018-9313-5}

Harvey, E., Koenecke, A., \& Kizilcec, R. F. (2025). ``Don't forget the teachers'': Towards an educator-centered understanding of harms from large language models in education. In \textit{Proceedings of the 2025 CHI Conference on Human Factors in Computing Systems} (pp. 1--19). ACM. \url{https://doi.org/10.1145/3706598.3713210}

Hedderich, M. A., Bazarova, N. N., Zou, W., Shim, R., Ma, X., \& Yang, Q. (2024). A piece of theatre: Investigating how teachers design LLM chatbots to assist adolescent cyberbullying education. In \textit{Proceedings of the 2024 CHI Conference on Human Factors in Computing Systems} (pp. 1--17). ACM. \url{https://doi.org/10.1145/3613904.3642379}

Hirata, Y., \& Hirata, Y. (2025). How do students' social and educational norms and practices affect GenAI chatbot-facilitated conversations? \textit{SN Computer Science}, \textit{6}(4), Article 325. \url{https://doi.org/10.1007/s42979-025-03826-4}

Hofstede, G. (1980). \textit{Culture's consequences: International differences in work-related values}. Sage.

Höök, K. (2000). Steps to take before intelligent user interfaces become real. \textit{Interacting with Computers}, \textit{12}(4), 409--426. \url{https://doi.org/10.1016/S0953-5438(99)00006-5}

Hou, X., Forsyth, C., Andrews-Todd, J., Rice, J., Cai, Z., Jiang, Y., Zapata-Rivera, D., \& Graesser, A. (2025). An LLM-enhanced multi-agent architecture for conversation-based assessment. In \textit{Artificial Intelligence in Education} (Lecture Notes in Computer Science, pp. 119--134). Springer. \url{https://doi.org/10.1007/978-3-031-98417-4_9}

Howie, S., Venter, E., \& van Staden, S. (2008). The effect of multilingual policies on performance and progression in reading literacy in South African primary schools. \textit{Educational Research and Evaluation}, \textit{14}(6), 551--560. \url{https://doi.org/10.1080/13803610802576775}

Joshi, P., Santy, S., Budhiraja, A., Bali, K., \& Choudhury, M. (2020). The state and fate of linguistic diversity and inclusion in the NLP world. In \textit{Proceedings of the 58th Annual Meeting of the Association for Computational Linguistics} (pp. 6282--6293). Association for Computational Linguistics. \url{https://doi.org/10.18653/v1/2020.acl-main.560}

Karpouzis, K., Pantazatos, D., Taouki, J., \& Meli, K. (2024). Tailoring education with GenAI: A new horizon in lesson planning. In \textit{2024 IEEE Global Engineering Education Conference (EDUCON)} (pp. 1--10). IEEE. \url{https://doi.org/10.1109/EDUCON60312.2024.10578690}

Kazemitabaar, M., Hou, X., Henley, A., Ericson, B. J., Weintrop, D., \& Grossman, T. (2023). How novices use LLM-based code generators to solve CS1 coding tasks in a self-paced learning environment. In \textit{Proceedings of the 23rd Koli Calling International Conference on Computing Education Research} (pp. 1--12). ACM. \url{https://doi.org/10.1145/3631802.3631806}

Khan Academy. (2023). \textit{Khanmigo: AI tutor by Khan Academy}. Retrieved May 6, 2025, from \url{https://www.khanacademy.org/khan-labs}

Klar, M. (2025). Using ChatGPT is easy, using it effectively is tough? A mixed methods study on K-12 students' perceptions, interaction patterns, and support for learning with generative AI chatbots. \textit{Smart Learning Environments}, \textit{12}(1), Article 32. \url{https://doi.org/10.1186/s40561-025-00385-2}

Kogut, B., \& Singh, H. (1988). The effect of national culture on the choice of entry mode. \textit{Journal of International Business Studies}, \textit{19}(3), 411--432. \url{https://doi.org/10.1057/palgrave.jibs.8490394}

Ladson-Billings, G. (1995). Toward a theory of culturally relevant pedagogy. \textit{American Educational Research Journal}, \textit{32}(3), 465--491. \url{https://doi.org/10.3102/00028312032003465}

Li, Y., Shirado, H., \& Das, S. (2025). Actions speak louder than words: Agent decisions reveal implicit biases in language models. In \textit{Proceedings of the 2025 ACM Conference on Fairness, Accountability, and Transparency} (pp. 3303--3325). ACM. \url{https://doi.org/10.1145/3715275.3732212}

Limke, A., Islam, S., Riahi, B., Tian, X., Hill, M., Catété, V., \& Barnes, T. (2025). What does it take to support problem solving in programming classrooms? A new framework from the K-12 teacher perspective. In \textit{Extended Abstracts of the CHI Conference on Human Factors in Computing Systems} (Article 591, pp. 1--7). ACM. \url{https://doi.org/10.1145/3706599.3719763}

Lin, X., Han, X., \& Qiu, J. (2025). Generative AI in special education: Teachers' insights on instructional enrichment vs. accommodations. In \textit{Extended Abstracts of the CHI Conference on Human Factors in Computing Systems} (pp. 1--6). ACM. \url{https://doi.org/10.1145/3706599.3719972}

Liu, Z. (2023). Cultural bias in large language models: A comprehensive analysis and mitigation strategies. \textit{Journal of Transcultural Communication}, \textit{3}(2), 224--244. \url{https://doi.org/10.1515/jtc-2023-0019}

Lu, J. G., Song, L. L., \& Zhang, L. D. (2025). Cultural tendencies in generative AI. \textit{Nature Human Behaviour}, \textit{9}(11), 2360--2369. \url{https://doi.org/10.1038/s41562-025-02242-1}

Lyu, H., Luo, J., Kang, J., \& Koenecke, A. (2025). Characterizing bias: Benchmarking large language models in Simplified versus Traditional Chinese. In \textit{Proceedings of the 2025 ACM Conference on Fairness, Accountability, and Transparency} (pp. 2815--2846). ACM. \url{https://doi.org/10.1145/3715275.3732182}

Ma, B. (2024). \textit{The impact of artificial intelligence tools on bilingual students in U.S. education: A study on academic language-learning, cultural sensitivity and inclusiveness} [Master's thesis, University of Washington]. ResearchWorks Archive. \url{https://hdl.handle.net/1773/51918}

Ma, S., \& Lei, L. (2024). The factors influencing teacher education students' willingness to adopt artificial intelligence technology for information-based teaching. \textit{Asia Pacific Journal of Education}, \textit{44}(1), 94--111. \url{https://doi.org/10.1080/02188791.2024.2305155}

MagicSchool AI. (2023). \textit{MagicSchool AI} [Computer software]. \url{https://www.magicschool.ai/}

Maslej, N., Fattorini, L., Perrault, R., Gil, Y., Parli, V., Kariuki, N., Capstick, E., Reuel, A., Brynjolfsson, E., Etchemendy, J., Ligett, K., Lyons, T., Manyika, J., Niebles, J. C., Shoham, Y., Wald, R., Walsh, T., Hamrah, A., Santarlasci, L., . . . Oak, S. (2025). \textit{Artificial Intelligence Index report 2025}. Stanford Institute for Human-Centered Artificial Intelligence. \url{https://hai.stanford.edu/ai-index/2025-ai-index-report}

Memon, T. D., \& Kwan, P. (2025). A collaborative model for integrating teacher and GenAI into future education. \textit{TechTrends}, \textit{69}(6), 1192--1206. \url{https://doi.org/10.1007/s11528-025-01105-w}

Merton, R. K. (1968). On sociological theories of the middle range. In \textit{Social theory and social structure} (Enlarged ed., pp. 39--72). Free Press.

Mhasakar, M., Baker-Ramos, R., Carter, B., Helekahi-Kaiwi, E.-B., \& Hester, J. (2025). ``I would never trust anything Western'': Kumu (educator) perspectives on use of LLMs for culturally revitalizing CS education in Hawaiian schools. In \textit{Extended Abstracts of the CHI Conference on Human Factors in Computing Systems} (pp. 1--10). ACM. \url{https://doi.org/10.1145/3706599.3720282}

Nistor, N., Göğüş, A., \& Lerche, T. (2013). Educational technology acceptance across national and professional cultures: A European study. \textit{Educational Technology Research and Development}, \textit{61}(4), 733--749. \url{https://doi.org/10.1007/s11423-013-9292-7}

OpenAI. (2022). \textit{ChatGPT} [Large language model]. \url{https://chat.openai.com/}

Paris, D. (2012). Culturally sustaining pedagogy: A needed change in stance, terminology, and practice. \textit{Educational Researcher}, \textit{41}(3), 93--97. \url{https://doi.org/10.3102/0013189X12441244}

Plata, M. (2008). Cultural sensitivity: The basis for culturally relevant teaching. \textit{Teacher Education and Practice}, \textit{21}(2), 181--200.

Reinecke, K., \& Bernstein, A. (2011). Improving performance, perceived usability, and aesthetics with culturally adaptive user interfaces. \textit{ACM Transactions on Computer-Human Interaction}, \textit{18}(2), Article 8. \url{https://doi.org/10.1145/1970378.1970382}

Reinecke, K., \& Bernstein, A. (2013). Knowing what a user likes: A design science approach to interfaces that automatically adapt to culture. \textit{MIS Quarterly}, \textit{37}(2), 427--453. \url{https://doi.org/10.25300/MISQ/2013/37.2.06}

Seekings, J., \& Nattrass, N. (2005). \textit{Class, race, and inequality in South Africa}. Yale University Press.

Shen, H., Knearem, T., Ghosh, R., Alkiek, K., Krishna, K., Liu, Y., Ma, Z., Petridis, S., Peng, Y.-H., Qiwei, L., Rakshit, S., Si, C., Xie, Y., Bigham, J. P., Bentley, F., Chai, J., Lipton, Z., Mei, Q., Mihalcea, R., . . . Jurgens, D. (2024). \textit{Towards bidirectional human-AI alignment: A systematic review for clarifications, framework, and future directions} (arXiv:2406.09264). arXiv. \url{https://doi.org/10.48550/arXiv.2406.09264}

Shen, H., Ma, Z., Ghosh, R., Knearem, T., Liu, M. X., Wu, T., Monroy-Hernández, A., Yang, D., Bosselut, A., Huang, F., Mitra, T., Chai, J., Hearst, M., Song, D., \& Li, Y. (2024). \textit{ICLR 2025 workshop on bidirectional human-AI alignment} [Workshop proposal]. ICLR 2025 Workshop Proposals, OpenReview. \url{https://openreview.net/forum?id=HcTiacDN8N}

Shenkar, O. (2001). Cultural distance revisited: Towards a more rigorous conceptualization and measurement of cultural differences. \textit{Journal of International Business Studies}, \textit{32}(3), 519--535. \url{https://doi.org/10.1057/palgrave.jibs.8490982}

Smith, A. E., Hudnut-Beumler, J., \& Scholer, S. J. (2017). Can discipline education be culturally sensitive? \textit{Maternal and Child Health Journal}, \textit{21}(1), 177--186. \url{https://doi.org/10.1007/s10995-016-2107-9}

Sousa, C. M. P., \& Bradley, F. (2008). Cultural distance and psychic distance: Refinements in conceptualisation and measurement. \textit{Journal of Marketing Management}, \textit{24}(5--6), 467--488. \url{https://doi.org/10.1362/026725708X325959}

Spaull, N., \& Jansen, J. D. (Eds.). (2019). \textit{South African schooling: The enigma of inequality}. Springer. \url{https://doi.org/10.1007/978-3-030-18811-5}

Statista. (2025). \textit{Countries with the highest income inequality 2025 (Gini index)} [Data set]. Retrieved July 16, 2025, from \url{https://www.statista.com/statistics/264627/ranking-of-the-20-countries-with-the-biggest-inequality-in-income-distribution/}

Straub, E. T. (2009). Understanding technology adoption: Theory and future directions for informal learning. \textit{Review of Educational Research}, \textit{79}(2), 625--649. \url{https://doi.org/10.3102/0034654308325896}

Tao, Y., Viberg, O., Baker, R. S., \& Kizilcec, R. F. (2024). Cultural bias and cultural alignment of large language models. \textit{PNAS Nexus}, \textit{3}(9), Article pgae346. \url{https://doi.org/10.1093/pnasnexus/pgae346}

Tu, M. Y.-P. (2025, April 25). \textit{AI, algorithms and Indigenous agency in Taiwan}. East Asia Forum. \url{https://eastasiaforum.org/2025/04/25/ai-algorithms-and-indigenous-agency-in-taiwan/}

Viberg, O., Cukurova, M., Feldman-Maggor, Y., Alexandron, G., Shirai, S., Kanemune, S., Wasson, B., Tømte, C., Spikol, D., Milrad, M., Coelho, R., \& Kizilcec, R. F. (2025). What explains teachers' trust in AI in education across six countries? \textit{International Journal of Artificial Intelligence in Education}, \textit{35}(3), 1288--1316. \url{https://doi.org/10.1007/s40593-024-00433-x}

Wang, D., Shan, D., Ju, R., Kao, B., Zhang, C., \& Chen, G. (2025). Investigating dialogic interaction in K12 online one-on-one mathematics tutoring using AI and sequence mining techniques. \textit{Education and Information Technologies}, \textit{30}(7), 9215--9240. \url{https://doi.org/10.1007/s10639-024-13195-9}

Wang, J., Xiao, R., Hou, X., Li, H., Tseng, Y. J., Stamper, J., \& Koedinger, K. (2025). LLMs to support K--12 teachers in culturally relevant pedagogy: An AI literacy example. In A. I. Cristea, E. Walker, Y. Lu, O. C. Santos, \& S. Isotani (Eds.), \textit{Artificial intelligence in education. Posters and late breaking results, workshops and tutorials, industry and innovation tracks, practitioners, doctoral consortium, blue sky, and WideAIED} (Communications in Computer and Information Science, Vol. 2591, pp. 152--160). Springer. \url{https://doi.org/10.1007/978-3-031-99264-3_19}

Woodruff, K., Hutson, J., \& Arnone, K. (2023). Perceptions and barriers to adopting artificial intelligence in K-12 education: A survey of educators in fifty states. In S. Mistretta (Ed.), \textit{Reimagining education: The role of e-learning, creativity, and technology in the post-pandemic era}. IntechOpen. \url{https://doi.org/10.5772/intechopen.1002741}

Wu, D., Zhang, X., Wang, K., Wu, L., \& Yang, W. (2025). A multi-level factors model affecting teachers' behavioral intention in AI-enabled education ecosystem. \textit{Educational Technology Research and Development}, \textit{73}(1), 135--167. \url{https://doi.org/10.1007/s11423-024-10419-0}

Xiao, Q., Fan, X., Simon, F. M., Zhang, B., \& Eslami, M. (2025). ``It might be technically impressive, but it's practically useless to us'': Motivations, practices, challenges, and opportunities for cross-functional collaboration around AI within the news industry. In \textit{Proceedings of the 2025 CHI Conference on Human Factors in Computing Systems} (pp. 1--19). ACM. \url{https://doi.org/10.1145/3706598.3714090}

Xiao, Q., Fan, X., Zhou, X., Su, Y., Lu, Z., Sap, M., \& Shen, H. (2026). How do users negotiate harmful value conflicts with AI companions? A study with Minion, a technology probe for in-situ human--AI conflict response. \textit{International Journal of Human--Computer Interaction}. Advance online publication. \url{https://doi.org/10.1080/10447318.2026.2740397}

Xiao, Q., Hu, Q., Xiao, J., Cao, H., \& Shen, H. (2026). Can GenAI move from individual use to collaborative work? Experiences, challenges, and opportunities of coordinating GenAI into collaborative newswork. In \textit{Proceedings of the 2026 CHI Conference on Human Factors in Computing Systems} (pp. 1--22). ACM. \url{https://doi.org/10.1145/3772318.3790984}

Xiao, R., Hou, X., Kumar, H., Moore, S., Stamper, J., \& Liut, M. (2024). A preliminary analysis of students' help requests with an LLM-powered chatbot when completing CS1 assignments. In Y. Shi, P. Brusilovsky, B. Akram, T. Price, J. Leinonen, K. Koedinger, \& A. Lan (Eds.), \textit{Proceedings of the 8th Educational Data Mining in Computer Science Education Workshop (CSEDM 2024)} (CEUR Workshop Proceedings, Vol. 3796). CEUR-WS.org. \url{https://ceur-ws.org/Vol-3796/CSEDM-24_paper_0046.pdf}

Yeo, A. (1996). Cultural user interfaces: A silver lining in cultural diversity. \textit{ACM SIGCHI Bulletin}, \textit{28}(3), 4--7. \url{https://doi.org/10.1145/231132.231133}

Yin, Y., Karumbaiah, S., Kim, Y., \& Saxena, D. (2026). When AI breaks, teachers repair: Pedagogical repair work in situated classroom practice. In \textit{Proceedings of the 2026 ACM Conference on Fairness, Accountability, and Transparency} (pp. 2206--2224). ACM. \url{https://doi.org/10.1145/3805689.3812281}

Zhang, P., \& Tur, G. (2024). A systematic review of ChatGPT use in K--12 education. \textit{European Journal of Education}, \textit{59}(2), Article e12599. \url{https://doi.org/10.1111/ejed.12599}

\end{hangparas}

\end{document}